\documentclass[12pt]{article}

\usepackage{fancyhdr}
\usepackage{isomath}
\usepackage{amsmath}
\usepackage{amsbsy}
\usepackage{amssymb}
\usepackage{amscd}
\usepackage{amsfonts}
\usepackage{graphics}
\usepackage{verbatim}
\usepackage{subfigure}
\usepackage{xspace}
\usepackage{euscript}
\usepackage{alltt}
\usepackage{boxedminipage}
\usepackage{float}
\usepackage{color}
\usepackage[,colorlinks,urlcolor=blue,citecolor=blue]{hyperref}
\usepackage[all]{xy}
\usepackage{t1enc}
\usepackage{times,exscale}
\usepackage{graphicx,calc}
\usepackage{mdwlist}
\usepackage{stmaryrd}
\usepackage{units}
\usepackage{setspace}
\usepackage{pdfpages}
\usepackage{hyperref}
\usepackage[usenames,dvipsnames]{xcolor}
\usepackage{cite}
\usepackage[normalem]{ulem}

\usepackage[english]{babel}
\usepackage[utf8]{inputenc}
\usepackage{algorithm}
\usepackage[noend]{algpseudocode}

\usepackage{CJKutf8}

\usepackage{isomath}
\usepackage{amsmath}
\usepackage{amssymb}
\usepackage{amscd}
\usepackage{amsfonts}

\newcommand{\R}{\mathbb R}

\newcommand{\beq}{\begin{equation}}
\newcommand{\eeq}{\end{equation}}
\newcommand{\beqs}{\begin{eqnarray}}
\newcommand{\eeqs}{\end{eqnarray}}
\newcommand{\half}{\frac{1}{2}}

\newcommand{\bfchi}{\mathbold {\chi}}

\newcommand{\bfsigma}{\mathbold {\sigma}}
\newcommand{\bfepsilon}{\mathbold {\epsilon}}

\newcommand{\bfomega}{\mathbold {\omega}}

\newcommand{\bfxi}{\mathbold {\xi}}
\newcommand{\bftau}{\mathbold {\tau}}

\newcommand{\bfzeta}{\mathbold {\zeta}}

\newcommand{\bfGamma}{\mathbold {\Gamma}}

\newcommand{\divergence}{\mathop{\rm div}\nolimits}

\newcommand{\parderiv}[2]{\frac{\partial #1}{\partial #2}}

\newcommand{\bfd}{{\mathbold d}}

\newcommand{\bff}{{\mathbold f}}
\newcommand{\bfg}{{\mathbold g}}

\newcommand{\bfk}{{\mathbold k}}

\newcommand{\bfp}{{\mathbold p}}
\newcommand{\bfq}{{\mathbold q}}

\newcommand{\bfs}{{\mathbold s}}

\newcommand{\bfu}{{\mathbold u}}
\newcommand{\bfv}{{\mathbold v}}
\newcommand{\bfw}{{\mathbold w}}
\newcommand{\bfx}{{\mathbold x}}
\newcommand{\bfy}{{\mathbold y}}
\newcommand{\bfz}{{\mathbold z}}

\newcommand{\bfC}{{\mathbold C}}
\newcommand{\bfD}{{\mathbold D}}
\newcommand{\bfE}{{\mathbold E}}
\newcommand{\bfF}{{\mathbold F}}

\newcommand{\bfH}{{\mathbold H}}
\newcommand{\bfI}{{\mathbold I}}

\newcommand{\bfP}{{\mathbold P}}
\newcommand{\bfQ}{{\mathbold Q}}

\newcommand{\bfT}{{\mathbold T}}

\newcommand{\bfZ}{{\mathbold Z}}

\graphicspath{ {imagedir/} }

\everymath{\displaystyle}

\usepackage[margin=20mm]{geometry}
\numberwithin{equation}{section}

\newcommand{\titlescript}{Effective Response of Heterogeneous Materials using the Recursive Projection Method}
\title{\titlescript}
\author{
	Xiaoyao Peng (\begin{CJK}{UTF8}{gbsn}彭潇瑶\end{CJK})\footnote{Email:\url{xiaoyaop@andrew.cmu.edu}}\ ,
	Dhriti Nepal\footnote{Email: \url{dhriti.nepal.1@us.af.mil}}\ , 
	Kaushik Dayal\footnote{Email: \url{Kaushik.Dayal@gmail.com}}
	\\ 
	\small{$^{\ast\ddagger}$Department of Civil and Environmental Engineering, Carnegie Mellon University}
	\\
	\small{$^\dagger$Composites Branch, Air Force Research Laboratory}
	\\
	\small{$^{\ddagger}$Center for Nonlinear Analysis, Carnegie Mellon University}
	\\
	\small{$^{\ddagger}$Department of Materials Science and Engineering, Carnegie Mellon University}
}
\date{\today}

\begin{document}

\pagestyle{fancyplain}
\lhead{\fancyplain
	{
		\scriptsize \titlescript \hfill Xiaoyao Peng, Dhriti Nepal, Kaushik Dayal \\ \textit{Comput. Methods. Appl. Mech. Engrg.} 364:112946, 2020. (\url{https://doi.org/10.1016/j.cma.2020.112946})
	}
	{
		\scriptsize \titlescript \hfill Xiaoyao Peng, Dhriti Nepal, Kaushik Dayal \\ \textit{Comput. Methods. Appl. Mech. Engrg.} 364:112946, 2020. (\url{https://doi.org/10.1016/j.cma.2020.112946})
	}
}
\rhead{\fancyplain{\ }{\ }}

\maketitle

\begin{abstract}
    This paper applies the Recursive Projection Method (RPM) to the problem of finding the effective mechanical response of a periodic heterogeneous solid.
    Previous works apply the Fast Fourier Transform (FFT) in combination with various fixed-point methods to solve the problem on the periodic unit cell.
    These have proven extremely powerful in a range of problems ranging from image-based modeling to dislocation plasticity.
    However, the fixed-point iterations can converge very slowly, or not at all, if the elastic properties have high contrast, such as in the case of voids.
    The paper examines the reasons for slow, or lack of convergence, in terms of a variational perspective.
    In particular, when the material contains regions with zero or very small stiffness, there is lack of uniqueness, and the energy landscape has flat or shallow directions.
    Therefore, in this work, the fixed-point iteration is replaced by the RPM iteration.
    The RPM uses the fixed-point iteration to adaptively identify the subspace on which fixed-point iterations are unstable, and performs Newton iterations only on the unstable subspace, while fixed-point iterations are performed on the complementary stable subspace.
    This combination of efficient fixed-point iterations where possible, and expensive but well-convergent Newton iterations where required, is shown to lead to robust and efficient convergence of the method.
    In particular, RPM-FFT converges well for a wide range of choices of the reference medium, while usual fixed-point iterations are usually sensitive to this choice.
\end{abstract}
\section{Introduction}
\label{sec:intro}

This paper proposes and applies the Recursive Projection Method (RPM) -- a method to adaptively combine Newton and fixed-point iterations -- to the problem of finding the effective mechanical response of a periodic heterogeneous linear elastic solid.
Specifically, given the average strain or stress, the goal is to compute the stress and strain distribution within the unit cell.
While the linear elastic setting is of interest in itself, it is also used extensively as part of the iterative solution process in nonlinear elastic or inelastic settings.

Early numerical works in the periodic setting exploited Fast Fourier Transforms (FFT)\footnote{Following common practice in this field, we denote also the class of algorithms that use Fast Fourier Transforms as FFT methods.} to partially simplify the problem; however, due to the heterogeneity of the elastic properties, the problem is not completely decoupled in Fourier space.
Therefore, fixed-point methods are used to complete the solution process \cite{moulinec1994fast, moulinec1998numerical, michel1999effective}.
This approach was used to compute the effective properties of heterogeneous materials that are linear, or nonlinear through iteration.
However, while the method is extremely fast for small elastic contrasts, the convergence is slow or even lost for higher elastic contrasts.
This rules out systems such as materials with voids, where the void is treated as a part of the body with zero stiffness, for instance.
This led to the formulation of the Accelerated FFT Method in \cite{michel2001computational}, based on a scheme proposed in \cite{eyre1999fast} for scalar conductivity problems, that modified the original method to handle high contrast composites.
A further refinement, proposed in \cite{monchiet2012polarization}, is the Polarization FFT Method, which has been shown to be a special case of a general class of methods in \cite{moulinec2014comparison}.
A separate promising approach is \cite{porta2013heterogeneity}, based on ideas from energy minimization, though they do not test it on the case of large elastic contrasts.

These methods have been applied in a variety of settings, including in highly nonlinear problems where they are used in each step of the iteration.
For instance, in the context of plasticity, we mention application at the macroscale \cite{lebensohn2001n,lebensohn2020spectral}, to continuous dislocation models \cite{beyerlein2016understanding,peng20203d}, as well as recent seminal work in coupling discrete dislocations with fast calculation of elastic interactions \cite{graham2016fast}.
Another important area of application is to phase-field models of microstructure formation in multifunctional materials, e.g. \cite{chen2013phase}.
The Fourier methods have also proven powerful in simulating experimental images of microstructure because of the ability to apply the method directly on image data without needing the identification of material boundaries and related challenges \cite{piazolo2015effect}.

The methods mentioned above -- starting from the original FFT method to the Accelerated and Polarization FFT methods -- are based on fixed-point methods or variations thereof; see \cite{moulinec2014comparison} for a summary.
Consequently, though some of these methods can converge for large elastic contrasts, the convergence gets slower as the contrast increases.
In addition, an important part of these methods is the notion of a homogeneous reference medium (for details, see Section \ref{sec:classical-FFT}).
The notion of a reference medium goes back to \cite{eshelby1959elastic}, and has proven to be a powerful concept in homogenization analysis \cite{agoras2013iterated,lipton1993inequalities}.
However, in the numerical setting, the convergence of the fixed-point methods can be very sensitive to the choice of the reference medium.
These reasons motivate the proposed work in applying the RPM to this problem.

A classical result is that the solution of the elasticity problem can be posed as a minimization of the strain energy density under standard assumptions, e.g. \cite{ciarlet1988three}.
That is, the balance of linear momentum, represented by $\divergence \bfsigma = {\bf 0}$ where $\bfsigma$ is the stress and $\bfy$ is the deformation, is equivalent to $\displaystyle \min_{\bfy} \int_\Omega W(\nabla\bfy)$, where $W$ is the elastic energy density, $\Omega$ is the domain, and $\displaystyle \parderiv{W}{\nabla\bfy} = \bfsigma$.
For simplicity, we have assumed displacement fixed on the entire boundary.
The perspective that the solution of the equilibrium equation is equivalent to the minimization of strain energy leads to useful insights into the solution methods.
In particular, consider the case of a heterogeneous elastic material with voids.
Typically, the voids are not taken to be part of the body $\Omega$, and instead they are defined through interior surfaces that are traction-free.
However, in the FFT method, one cannot easily treat the voids in this manner because the discretization must be uniform for efficiency; hence, the voids are taken to be part of $\Omega$, and treated as regions with zero stiffness. 
The zero stiffness regions imply that there is infinite elastic contrast, but also {\em loss of uniqueness} of the solution.

The uniqueness of solutions is an elementary calculation in linear elasticity, e.g. \cite{ciarlet1988three}.
Assuming a boundary value problem with 2 solutions $\bfy_1$ and $\bfy_2$, it can be shown that the difference $\bfu := \bfy_1 - \bfy_2$ satisfies $\displaystyle\int_\Omega \bfepsilon_\bfu:\bfC \bfepsilon_\bfu = 0$, where $\bfC$ is the linear elastic stiffness tensor, and $\bfepsilon_\bfu$ is the small strain associated with $\bfu$.
If $\bfC$ is pointwise positive-definite, then $\bfepsilon_\bfu \equiv 0$, and we have uniqueness up to rigid motions; Dirichlet boundary conditions on a finite part of the boundary fixes the rigid motion.

From a heuristic perspective, the lack of uniqueness of the solution in a material with zero-stiffness regions is reasonable: the fictitious material points in the void region can have arbitrary displacement (while respecting the required smoothness), and this does not cost any elastic energy or affect the physics of the problem.
Hence, we can expect soft or flat directions in energy landscape.
When we have large elastic contrast due to regions with very low stiffness, the energy will be shallow though perhaps not completely flat.
Convergence along these directions can be exceedingly slow, or even lost, in a fixed-point method.

We briefly note that for rigid inclusions, the strain within the rigid objects must vanish, while the stress can be arbitrary as long as smoothness and equilibrium are satisfied.  The discussion above can be readily adapted to this setting using the {\em complementary} strain energy density.

Given the expectation that the energy landscape is flat or shallow in some directions, the use of Newton or similar iterations is likely to provide convergence within a relatively small number of iterations.
However, each Newton iteration is far more expensive than a fixed-point iteration, and increasingly so as the dimension of the solution space increases.
The RPM, proposed in \cite{shroff1993stabilization}, provides the possibility of a balance between the fixed-point and the Newton methods.
The first key aspect of RPM is that it uses the fixed-point iterations to adaptively identify the unstable subspace.
That is, starting from the assumption that fixed-point methods will converge when applied to the problem, RPM examines successive outcomes from the fixed-point iteration scheme to discern if, and how, the basis of the unstable subspace should be augmented.
The second key aspect of RPM is that as the unstable subspace is built up, Newton iterations are performed on the unstable subspace.
This enables a balance between the less expensive but non-converging fixed-point method and the more expensive but better converging Newton method.

We point out that while Newton iterations are performed only on the unstable subspace, the fixed-point iterations are performed on the {\em entire} space.
The reason to perform fixed-point iterations not only on the stable subspace, but rather on the entire subspace which is slightly more expensive, is that it enables the use of existing fixed-point methods with minimal changes to existing code and algorithm.
In fact, we exploit this feature to apply RPM to both the original FFT method \cite{moulinec1998numerical} as well as to the Accelerated FFT method \cite{moulinec2014comparison}.

Briefly, the key outcomes of our comparisons of RPM-FFT with the existing methods is that the RPM-FFT method is faster than the other methods for large elastic contrasts.
However, an even more important advantage appears to be that the convergence of RPM-FFT is very robust with respect to the choice of the reference homogeneous material, whereas the fixed-point methods are sensitive to this choice; this is particularly important since there are only some heuristic ideas about how to optimally choose the reference material.

For brevity, we refer below to the method proposed in this paper as {\em RPM-FFT}; the conventional algorithm from \cite{moulinec1998numerical} as {\em Classical FFT}; the method proposed in \cite{michel2001computational} as {\em Accelerated FFT}; and the method proposed in \cite{monchiet2012polarization} as {\em Polarization FFT}.

The manuscript is organized as follows:
\begin{itemize}
    \item Section \ref{sec:classical-FFT} summarizes the Classical FFT method;
    \item Section \ref{sec:RPM-FFT} describes the RPM-FFT method;
    \item Section \ref{sec:variation} provides a variational perspective;
    \item Section \ref{sec:numerics} provides numerical examples and comparisons.
\end{itemize}

\section{Classical FFT Method Based on Fixed-Point Iteration}
\label{sec:classical-FFT}

The original FFT method proposed by Moulinec and Suquet \cite{moulinec1998numerical}, and some subsequent improvements, are briefly summarized below in the setting of linear elasticity.
The goal is to solve the linear elasticity problem on a unit cell $V$ in a periodic geometry with a heterogeneous medium.
The average symmetric strain tensor $\bfE$ is given.

We use $\mathcal{F}[\cdot], \hat{\cdot}, \mathcal{F}^{-1}[\cdot]$ respectively to denote the Fourier transform, Fourier space representation, and inverse Fourier transform.

The displacement field $\bfu(\bfx)$ is decomposed into a linear part and a zero-mean fluctuating part: $\bfu(\bfx) = \bfE \cdot \bfx + \bfv(\bfx)$, ignoring a constant.
This implies the strain decomposition $\bfepsilon_\bfu(\bfx) = \bfE + \bfepsilon_\bfv(\bfx)$, where $\bfepsilon_{(\cdot)} := \half\left(\nabla(\cdot) + \nabla(\cdot)^T\right)$ is the strain.

We notice that $\hat{\bfepsilon}_\bfv (\bfk) = \hat{\bfepsilon}_\bfu (\bfk) \ \forall \ \bfk \neq \bf0$, and $\hat{\bfepsilon}_\bfu (\bf0) = \bfE$ while $\hat{\bfepsilon}_{\bfv}(\bf 0) = 0$.
Therefore, for conciseness after this section, we will use $\hat{\bfepsilon}$ to represent both $\hat{\bfepsilon}_\bfv$ and $\hat{\bfepsilon}_\bfu$, and $\bfepsilon$ to represent both $\bfepsilon_\bfu$ and $\bfepsilon_\bfv$.

The unit cell problem can then be written as  
\begin{align}
\label{eqn:local}
	\divergence \bfsigma(\bfx) & = {\bf 0} \quad \forall \ \bfx \in V
	\\
	\bfsigma(\bfx) &= \bfC(\bfx) \bfepsilon_\bfu(\bfx)
\end{align}
where $\bfsigma$ is the stress field, and is periodic.

We can then introduce, following \cite{monchiet2012polarization,moulinec2014comparison,eyre1999fast} and others, a homogeneous linear elastic comparison medium with stiffness $\bfC^0$ and the polarization $\bftau(\bfx) := \left(\bfC(\bfx) - \bfC^0 \right) \bfepsilon_\bfu(\bfx)$.
This enables us to rewrite the stress-strain relation as:
\begin{equation}
	\bfsigma(\bfx)=\bfC^0 \bfepsilon_\bfv(\bfx) + \bftau(\bfx) + \bfC^0 \bfE
\end{equation}

The solution to the unit cell problem can be written in terms of the periodic Greens function $\bfGamma^0$ corresponding to $\bfC^0$:
\begin{equation}
\label{eqn:local_real}
	\bfepsilon_\bfv (\bfx) = -\bfGamma^0(\bfx)*\bftau(\bfx) \quad \forall \ \bfx \in V
\end{equation}
The Fourier space representation of this relation is:
\begin{eqnarray}
\label{eqn:local_fourier}
	\hat{\bfepsilon}_\bfv (\bfk)=-\hat{\bfGamma}^0(\bfk) \hat{\bftau}(\bfk) \quad \forall \ \bfk \neq \bf0; \ \hat{\bfepsilon}_\bfv (\bf0) = \bf0
\end{eqnarray}

Typical FFT methods use the fact that taking the (fast) Fourier transform and then multiplying in Fourier space is much faster than convolution in real-space.
However, we note that both the real and Fourier representations in  (\ref{eqn:local_real}, \ref{eqn:local_fourier}) are implicit, through the dependence of the effective forcing $\bftau$ on the solution $\bfu$.
Therefore, further work is required to solve this (linear) implicit equation.

From a linear algebra perspective, the Fourier transform takes us to the eigenbasis in a homogeneous problem; in a heterogeneous problem, the Fourier transform brings us close to the eigenbasis, and further calculations are required to completely diagonalize the operator.
Roughly, one can make an analogy between the effort to completely diagonalize and the effort to solve the implicit equation in Fourier space.

The implicit equations are typically solved with fixed point methods, and variations of these.
In \cite{moulinec1998numerical}, for instance, they use the iteration $\hat{\bfepsilon}^{i+1}(\bfk)=\hat{\bfepsilon}^i(\bfk)-\hat{\bfGamma}^0(\bfk)\hat{\bfsigma}^i(\bfk)$; see Algorithm \ref{alg:FFT}.
Like all fixed point methods, convergence is slow or lost when the problem contains an unstable subspace.
This is precisely the case when there is large contrast leading to flat energy landscapes and loss of uniqueness, as has been observed in practice \cite{eyre1999fast}.


\begin{algorithm}
\caption{Classical FFT algorithm \cite{moulinec1998numerical}}\label{alg:FFT}
\begin{algorithmic}[1]

\State \textbf{{Initialization} } 

\State $\bfepsilon^0 \gets \bfE$; \quad $\bfsigma^0 \gets \bfC(\bfx):\bfepsilon^0$

\State \textbf{Iteration} ($\bfepsilon^0, \bfsigma^0$)

\While{$\text{error} > \text{tolerance}$}

	\State $\hat{\bfepsilon}^i \gets \mathcal{F}[\bfepsilon^i]; \quad \hat{\bfsigma}^i \gets \mathcal{F}[\bfsigma^i]$  	\Comment{FFT on $\bfepsilon$ and $\bfsigma$}
	
	\State $\hat{\bfepsilon}^{i+1} \gets \hat{\bfepsilon}^i-\hat{\bfGamma}^0:\hat{\bfsigma}^i$ \Comment{$\forall \ \bfk \neq \mathbf{0}; \hat{\bfepsilon}^{i+1}(\mathbf{0}) = \bfE$}
	
	\State $\bfepsilon^{i+1} \gets \mathcal{F}^{-1}[\hat{\bfepsilon}^{i+1}]$ 
	
	\State $\bfsigma^{i+1} \gets \bfC(\bfx):\bfepsilon^{i+1}$
    
    \State $\text{error} \gets \frac{ (\langle {\left\lVert \bfk \cdot \hat{\bfsigma}(\bfk) \right\rVert}^2 \rangle)^{1/2} }{ \left\lVert \hat{\bfsigma}(\mathbf{0}) \right\rVert} \equiv \frac{ (\langle {\left\lVert \divergence \bfsigma \right\rVert}^2 \rangle)^{1/2} }{ \left\lVert \langle\bfsigma\rangle \right\rVert}$
\EndWhile
\State \textbf{end while}
\end{algorithmic}
\end{algorithm}



\section{The Proposed RPM-FFT Method}
\label{sec:RPM-FFT}

We first outline the key elements of the recursive projection method following the seminal work of Shroff and Keller \cite{shroff1993stabilization}, and then the application of RPM to the FFT approach.


When using fixed-point iterative procedures to solve nonlinear problems, slow convergence or even divergence can be caused by a small number of eigenvalues of the Jacobian matrix approaching or leaving the unit disk $\{|z|<1\}$. 
The Recursive Projection Method (RPM) is a stabilization procedure that can recover the convergence.

Consider a problem in $N$ dimensions posed on $\R^N$.
The method begins by dividing the space $\mathbb{R}^N$ into $\mathbb{P}$, that we denote as the {\em unstable eigenspace}, and its orthogonal complement $\mathbb{Q}$.
$\mathbb{P}$ and $\mathbb{Q}$ correspond respectively to the eigenspace of those eigenvalues leaving the disk and those inside the disk. 
Then, the fixed-point iteration is performed only on the complement $\mathbb{Q}$ while Newton's method -- or, in practice, a variation such as quasi-Newton -- is performed on the unstable eigenspace $\mathbb{P}$.

Writing the original problem to be solved in the fixed-point form as follows:
\begin{equation}
\label{eq:potype}
	\bfu = \bfF(\bfu),\quad \bfF{:} \ \mathbb{R}^N \to \mathbb{R}^N
\end{equation}
the fixed-point iteration can be written:
\begin{equation}
	\bfu^{(\nu+1)} = \bfF(\bfu^{(\nu)})
\end{equation}
where $\nu$ indicates the iteration number.
While the eigenvalues of the Jacobian matrix $\bfF_\bfu := \parderiv{\bfF}{\bfu}$ lie inside of the disk
\begin{equation}
	K_\delta= \{|z|\leq 1-\delta\} \quad \text{for some} \, \delta>0
\end{equation}
and the initial value $\bfu^{(0)}$ is sufficiently close to the solution, the fixed-point iteration will converge.
A small positive constant $\delta$ is used to ensure that one is sufficiently far from the boundary between stable and unstable fixed point iterations.


Suppose now that $m$ eigenvalues lie outside the disk $K_\delta$
\begin{equation}
	|\mu_1| \geq \cdots \geq |\mu_m| > 1-\delta \geq |\mu_{m+1}|\geq \cdots \geq |\mu_N|
\end{equation}
Let $\bfZ\in \mathbb{R}^{N \times m} $ be the orthonormal basis for $\mathbb{P}$.
The projectors of the subspaces $\mathbb{P}$ and $\mathbb{Q}$ are respectively $\bfP=\bfZ\bfZ^{T}$ and $\bfQ=\bfI-\bfZ\bfZ^{T}$. 
For each $\bfu\in\mathbb{R}^n$, there is a unique decomposition:
\begin{equation}
	\bfu
	=
	\bfp+\bfq,\quad 
	\bfp\equiv \bfP \bfu\in \mathbb{P},\quad 
	\bfq\equiv \bfQ \bfu\in\mathbb{Q}
\end{equation}
This allows the introduction of $\bff(\bfp,\bfq) := \bfP \bfF(\bfp+\bfq)$ and $\bfg(\bfp,\bfq) := \bfQ \bfF(\bfp+\bfq)$.

Applying Newton's method on the unstable subspace $\mathbb{P}$:
\begin{equation}
\label{qunt}
	\bfp^{(\nu+1)}=\bfp^{(\nu)}+(\bfI-\bff_\bfp^{(\nu)})^{-1}(\bff(\bfp^{(\nu)},\bfq^{(\nu)})-\bfp^{(\nu)})
\end{equation} 
where $\bff_\bfp(\bfp^{(\nu)},\bfq^{(\nu)}) = \bfP\bfF_u(\bfu^{(\nu)})\bfP$ is the derivative of $\bff$ with respect to $\bfp$.

We continue to use fixed-point iteration on $\mathbb{Q}$:
\begin{equation}
\label{eq:fp}
	\bfq^{(\nu+1)}=\bfg(\bfp^{(\nu)},\bfq^{(\nu)})
\end{equation}

Efficiently identifying the unstable eigenspace $\mathbb{P}$ is an essential feature of RPM. 
This corresponds to finding the basis $\bfZ$; or, in practice, an approximation of $\bfZ$ since the Jacobian matrix can be expensive to compute. 
In \cite{shroff1993stabilization}, this is accomplished directly by monitoring the convergence rate of $\bfq^{(\nu)}$, without computing the Jacobian matrix. 
This is an important feature that significantly reduces the computational cost, and we follow that procedure here.



The method begins by assuming that the unstable eigenspace $\mathbb{P}$ is $0$-dimensional and $\bfZ = {\bf 0}$. 
We then iteratively build up the unstable eigenspace by adding to $\bfZ$ the eigenspace corresponding to those eigenvalues that approach the unit disk . 
The eigenspace that corresponds to the eigenvalues approaching the unit disk corresponds to the dominant eigenspace of $\bfg_q=\bfQ\bfF_u\bfQ$.

If the fixed-point iteration does not converge within the specified limit of iterations $n_{max}$, the eigenspace corresponding to the eigenvalues approaching the unit disk is identified as follows.
In \cite{shroff1993stabilization}, it is shown that $\Delta\bfq^{(\nu)}\equiv\bfq^{(\nu+1)}-\bfq^{(\nu)}\approx(\bfg_q)^\nu\Delta\bfq^{(0)}$, i.e., a power iteration with $\bfg_q$ applied to $\Delta\bfq^{(0)}$. 
From the properties of power iterations, $\Delta\bfq^{(\nu)}$ identifies the eigenspace of $\bfg_q$ that corresponds to the largest eigenvalue, assuming that $\Delta\bfq^{(0)}$ has a nonzero component in this space.

Following \cite{shroff1993stabilization}, we approximate the dominant eigenspace of $\bfg_q$ using QR factorization to write:
\begin{equation}
    \bfD\equiv\left[\Delta\bfq^{(\nu)},\Delta\bfq^{(\nu-1)} \right] =: \tilde{\bfD}\bfT
\end{equation}
where $\tilde{\bfD} \in \mathbb{R}^{N\times2}$ is orthogonal and $\bfT \in \mathbb{R}^{2\times2}$ is upper triangular. 
When $\bfT_{11}\gg\bfT_{22}$, we add the first column of $\tilde{\bfD}$ to the basis $\bfZ$, corresponding to one real eigenvalue approaching the unit disk.
Else, we add both columns of $\tilde{\bfD}$ to the basis $\bfZ$, corresponding to a complex conjugate pair of eigenvalues approaching the unit disk. 
Alternate approximations are further discussed in \cite{shroff1993stabilization}.   

To apply the RPM method to the specific context of the FFT fixed-point algorithm, we formally define the fixed-point operator $\bfF$ from Algorithm \ref{alg:FFT} through the equation:
\begin{equation}
    \bfepsilon^{i+1} 
    = 
    \mathcal{F}^{-1}\left[ \mathcal{F}[\bfepsilon^i] - \hat{\Gamma^0}:\mathcal{F}[\bfC(\bfx):\bfepsilon^i]\right]
\end{equation}

Define $\bfz := \bfZ^T\bfp\in \mathbb{R}^m$ to represent $\bfp$ in the basis $\bfZ$. 
Then the stable / unstable decomposition can also be written as:
\begin{equation}
	\bfu=\bfZ\bfz+\bfq,\quad \bfp=\bfZ\bfz,\quad \bfq=(\bfI-\bfZ\bfZ^T)\bfu
\end{equation}  
It is suggested in \cite{shroff1993stabilization} that it is often sufficient for convergence to simply compute only once the quantity $(\bfI-\bff_p)^{-1}$ using finite differences.
In our calculations in this paper, we have also found this to be sufficient for convergence, and of course very efficient.

Defining $\bfH := \bfZ^T \bfF_\bfu \bfZ$, the iteration (\ref{qunt}, \ref{eq:fp}) can now be rewritten in a form that is more efficient and transparent for implementation:
\begin{align}
	\bfz^{(\nu+1)} &=
	    \bfz^{(\nu)}+(\bfI-\bfH)^{-1}(\bfZ^T\bfF(\bfu^{(\nu)})-\bfz^{(\nu)})
    \\
    \bfq^{(\nu+1)} &= 
        (\bfI-\bfZ\bfZ^T)\bfF(\bfu^{(\nu)})
    \\
	\bfu^{(\nu+1)} & =
	    \bfZ\bfz^{(\nu+1)}+\bfq^{(\nu+1)}
\end{align}
Finally, integrating the FFT method into RPM, gives us Algorithm \ref{alg:RPM}.

\begin{algorithm}
\caption{RPM-FFT algorithm}
\label{alg:RPM}
\begin{algorithmic}[1]
\State \textbf{{Initialization} } 
\State $\bfepsilon^0 \gets \bfE$; $\bfxi^0 \gets \bfF(\bfepsilon^0)$

\State \textbf{Iteration} ($\bfepsilon^0, \bfxi^0$)
\While{error $>$ tolerance}
	
	\State $\bfz^{(\nu)} \gets \bfZ^T\bfepsilon^{(\nu)}$;  $\bfzeta^{(\nu)} \gets \bfZ^T \bfxi^{(\nu)}$ \Comment{Project to unstable subspace}
	
	\State $\bfz^{(\nu+1)} \gets \bfz^{(\nu)}+(\bfI-\bfH)^{-1} (\bfzeta^{(\nu)} - \bfz^{(\nu)})$ \Comment{Newton iteration}
	
	\State $\bfq^{(\nu+1)} \gets \bfxi^{(\nu)}-\bfZ\bfzeta^{(\nu)}$
	\State $\bfepsilon^{(\nu+1)} \gets \bfZ\bfz^{(\nu+1)}+\bfq^{(\nu+1)}$
	\State $\bfxi^{(\nu+1)} \gets \bfF(\bfepsilon^{(\nu+1)})$ \Comment{Fixed-point iteration}
	\State $\nu=\nu+1$
	\If{$\nu>n_{max}$}  \Comment{Start Newton iteration if number of fixed-point iterations exceeds $n_{max}$}
    	\State $\bfZ \gets \mathbf{In}(\bfq,\bfZ)$ \Comment{Increase basis size}
        \State $\bfH \gets \bfF_u\bfZ$ \Comment{Compute additional column of $\bfH$}
        \State $\nu=0$
	\EndIf
    \State \textbf{end if}
    \State error $\gets \frac{ (\langle {\left\lVert \bfk \cdot \hat{\bfsigma}(\bfk) \right\rVert}^2 \rangle)^{1/2} }{ \left\lVert \hat{\bfsigma}(\mathbf{0}) \right\rVert}$
\EndWhile
\State \textbf{end while}
\end{algorithmic}
\end{algorithm}

An important feature is that the fixed-point iteration is conducted on the entire solution space and not only on the stable subspace.
After the fixed-point iteration, only that part of the outcome corresponding to the stable part is retained in constructing the next iterate of the approximate solution.
While marginally more computationally expensive than using a fixed-point iteration only on the stable subspace, it has the important advantage that standard existing fixed-point codes can directly be used with minimal modification as the RPM ``wraps around'' the fixed-point method.


\section{A Variational Perspective}
\label{sec:variation}

We begin with noting an analogy between the problem of interest here and the energy-minimization formulation of electrostatic fields in matter.
The field equations and constitutive response of electrostatic fields in matter are respectively:
\begin{equation}
    \divergence \nabla \phi = \divergence \bfp \text{ on } \Omega, \quad \bfp(\bfx) = \tilde{\bfp}(\nabla\phi(\bfx),\bfx), \quad \text{ BCs: } \phi(\bfx) = V_0 (\bfx) \ \forall \bfx \in \partial\Omega
\end{equation}
where $\phi$ is the electric potential, $\bfp$ is the dipole per unit volume induced by the electric field $\nabla \phi$, and $\tilde{\bfp}$ is the material-dependent response function that can be an explicit function of position.
For simplicity, we have assumed Dirichlet boundary conditions $V_0$ on the boundary $\partial \Omega$ of the domain $\Omega$, but it is straightforward to use more general BCs.
Note that we neglected various constants such as the permittivity of vacuum.

Under some physically-motivated conditions on $\tilde{\bfp}$, this can be reformulated as the minimization of electrostatic energy.
For instance, we consider the case of a linear dielectric which is analogous to the linear elastic medium considered in this paper, in which case $\tilde{\bfp}$ takes the form:
\begin{equation}
    \tilde{\bfp}(\bfE(\bfx),\bfx) = \bfchi(\bfx) \bfE
\end{equation}
where $\bfchi(\bfx)$ is the second-order electrical susceptibility tensor.

Following \cite{james1990frustration} for magnetostatics  -- see also \cite{shu2001domain} and \cite{yang2011completely} for electrostatics -- we can write this as an energy minimization with a nonlocal constraint:
\begin{equation}
    \min E_{electro}[\bfp,\phi], \text{ with } E_{electro}[\bfp,\phi] = \half \int_{\Omega} \left( \bfp \cdot \bfchi^{-1} \bfp + |\nabla\phi|^2 \right), \text{ subject to } \divergence \nabla \phi = \divergence \bfp \text{ on } \Omega
\end{equation}
with BCs $\phi(\bfx) = V_0 (\bfx) \ \forall \bfx \in \partial\Omega$.

From the fundamental physics of electrostatics problems, related to the motion of charges under an electric field, we have $\bfchi$ is positive-definite pointwise.
Therefore, from an elementary application of the direct method of the calculus of variations, a unique minimizer exists; loosely, the energy is bounded below and grows quadratically in all directions from the positive-definiteness of $\bfchi$.
The nonlocal constraint, while somewhat nonstandard, can be readily dealt with, e.g. \cite{james1990frustration}.


We turn now to linear inhomogeneous elasticity.
We recall the balance of linear momentum:
\begin{equation}
    \divergence \bfC(\bfx) \bfepsilon_\bfu = 0
\end{equation}
posed on the periodic unit cell $V$ with periodic BCs.
We decompose $\bfC(\bfx) = \bfC_0 + \bfd(\bfx)$, where $\bfC_0$ is the homogeneous reference medium.
This gives:
\begin{equation}
    \divergence \bfC_0 \bfepsilon_\bfu = - \divergence \bfd \bfepsilon_\bfu
\end{equation}

Recall from Section \ref{sec:classical-FFT} that we defined the polarization field $\bftau := \bfd \bfepsilon_\bfu$.
Following closely the electrostatic case, we can pose this as a nonlocally-constrained critical-point problem.
Define the energy $E$:
\begin{equation}
     E[\bftau,\bfu] =  \half \int_{V} \left( \bftau:\bfd^{-1}\bftau + \bfepsilon_\bfu : \bfC_0 \bfepsilon_\bfu \right), \text{ subject to } \divergence \bfC_0 \bfepsilon_\bfu = - \divergence \bftau
\end{equation}

Use the variation $\bftau \to \bftau + \epsilon \bfomega$ and $\bfu \to \bfu + \epsilon\bfw$:
\begin{equation}
\label{eqn:variation}
    \delta E = \int_{V} \left( \bfomega : \bfd^{-1}{\bftau} + \bfepsilon_\bfu : \bfC_0 \bfepsilon_\bfw \right)
\end{equation}
The variations $\bfomega$ and $\bfw$ are not independent; they are constrained to satisfy:
\begin{equation}
    \divergence\bfC_0 \bfepsilon_\bfw = -\divergence \bfomega
\end{equation}
On both sides of the equation above, do the following: (1) take the inner product with $\bfepsilon_\bfu$; (2) integrate over $V$; (3) use integration-by-parts to move the derivatives.  The result of these operations is:
\begin{equation}
\label{eqn:constrained-variation}
    \int_{V} \bfepsilon_\bfu : \bfC_0 \bfepsilon_\bfw 
    =
    \int_{V} - \bfomega : \bfepsilon_\bfu
\end{equation}
where the boundary terms cancel out due to periodicity.

Using (\ref{eqn:constrained-variation}) in (\ref{eqn:variation}), and requiring $\delta E = 0$ for arbitrary $\bfomega$, we have:
\begin{equation}
    \bfd^{-1}\bftau - \bfepsilon_\bfu = 0, \quad \divergence \bfC_0 \bfepsilon_\bfu = - \divergence \bftau 
\end{equation}

We highlight here that the tensor $\bfd(\bfx)$ is generally not positive-definite pointwise, particularly in the important case of high elastic contrasts.
For instance, consider a material with voids, where the non-voided material is stable, i.e. $\bfC(\bfx)$ is positive-definite in the non-voided material, and $\bfC(\bfx) = \bf0$ in the void region.
The reference medium is taken to be stable, i.e. $\bfC_0$ is positive-definite to enable solution of the constraint equation.
Then, $\bfd = -\bfC_0$ in the voids, and is not positive-definite pointwise.
Therefore, in general, we do not expect existence of minimizers for $E$.

Loosely, the energy has unstable directions which are not bounded below.
Therefore, gradient descent methods -- which can be written as fixed-point methods as shown below for this energy formulation and by \cite{schneider2017fft} -- do not converge.
On the other hand, using Newton iterations along the unstable directions -- once these are identified -- can greatly improve convergence, which is precisely the role of the Recursive Projection Method.

\subsection{Fixed-Point Iterations as Gradient Descent}

We consider a gradient / steepest descent approach to minimizing the energy $E$:
\begin{equation}
	E = \half \int_{\Omega_\#} \left(\bftau : \bfd^{-1} \bftau + \half \bfepsilon_\bfu : \bfC_0 \bfepsilon_\bfu \right), \text{ subject to } \divergence \bfC_0 \bfepsilon_\bfu = \divergence \bftau
\end{equation}
The gradient direction in function space is given by $\bfd^{-1}\bftau - \bfepsilon_\bfu$, and a gradient flow in the standard $L^2$ norm is $\dot{\bftau} \sim  \bfd^{-1}\bftau - \bfepsilon_\bfu$.

Consider an explicit update scheme where the constraint equation is also updated at each iteration.
We obtain the following fixed-point algorithm:
\begin{equation}
\begin{split}
	& 1. \text{ given } \bfepsilon_\bfu^i \\
	& 2. \ \bftau^{i+1} = \bftau^i + a (\bfd^{-1}\bftau^i - \bfepsilon_\bfu^i) \\
	& 3. \divergence \bfC_0 \bfepsilon_\bfu^{i+1} = \divergence \bftau^{i+1} \\
	& 4. \text{ loop }
\end{split}
\end{equation}
where $a$ is related to the fictitious timestep and mobility, and the superscript indexes the iterations.
Notice that the constraint update (step 3 above) requires solution of a homogeneous periodic linear elastic problem with a given right side.
This can be done very efficiently using fast Fourier transforms.
Each iteration in the loop above is therefore extremely quick, but the key question is whether the algorithm converges and, if so, how many iterations it takes.
Given that $E$ may not possess minimizers, it follows that gradient descent may not converge.

The Classical FFT, summarized in \cite{moulinec2014comparison}, can be written in the fixed point form:
\begin{equation}
\begin{split}
	& 1. \text{ given } \bfepsilon_\bfu^i \\
	& 2. \ \bftau^{i+1} = \bfd \bfepsilon_\bfu^i \\
	& 3. \divergence \bfC_0 \bfepsilon_\bfu^{i+1} = \divergence \bftau^{i+1} \\
	& 4. \text{ loop }
\end{split}
\end{equation}
This can be considered as derived from the explicit update scheme above, except that the update for $\bftau$ in step 2 goes directly to the minimum -- for a given $\bfepsilon_\bfu^i$ --  using that the energy is quadratic.
As in the explicit update scheme, this is very fast for each iteration, but whether it converges, and the rate of convergence, depend on the structure of the energy landscape.

The Polarization FFT proposed by \cite{monchiet2012polarization} can be rewritten in the form:
\begin{equation}
\begin{split}
	& 1. \text{ given } \bfepsilon_\bfu^i \\
	& 2. \ \bftau^{i+1} = \bfd \bfepsilon_\bfu^i \\
	& 3. \divergence \bfC_0 \bfepsilon_\bfu^{i+1} 
	    = 
	    \divergence \bfC_0 \bfd^{-1} \bftau^{i+1} 
	    + \beta \divergence \bfC_0 \bfd^{-1} \bfC_0 \bfepsilon_\bfu^i 
	    - \divergence \bfC_0 \bfd^{-1} \bfC_0 \tilde{\bfepsilon}_\bfu^i \\
	& \qquad \text{ where } \tilde{\bfepsilon}_\bfu^i \text{ solves } 
	    \divergence \bfC_0 \tilde{\bfepsilon}_\bfu^i = \divergence \left( \alpha \bftau^i + (\alpha+\beta) \bfC_0 \bfepsilon_\bfu^i \right) \\
	& 4. \text{ loop }
\end{split}
\end{equation}
While this form is not as transparent or convenient, it enables us to compare the methods in the context of gradient descent.
We notice that in this form, the polarization method can be considered as a relaxation method that mixes in the value from the previous iteration, with $\alpha$ and $\beta$ being the relaxation parameters.


\section{Numerical Comparisons Between Classical FFT, Accelerated FFT, and RPM-FFT} 
\label{sec:numerics}

\begin{figure}[t!]
	\begin{center}
		\fbox{
			\begin{minipage}{165mm}
				\begin{center}
                    \includegraphics[width=80mm]{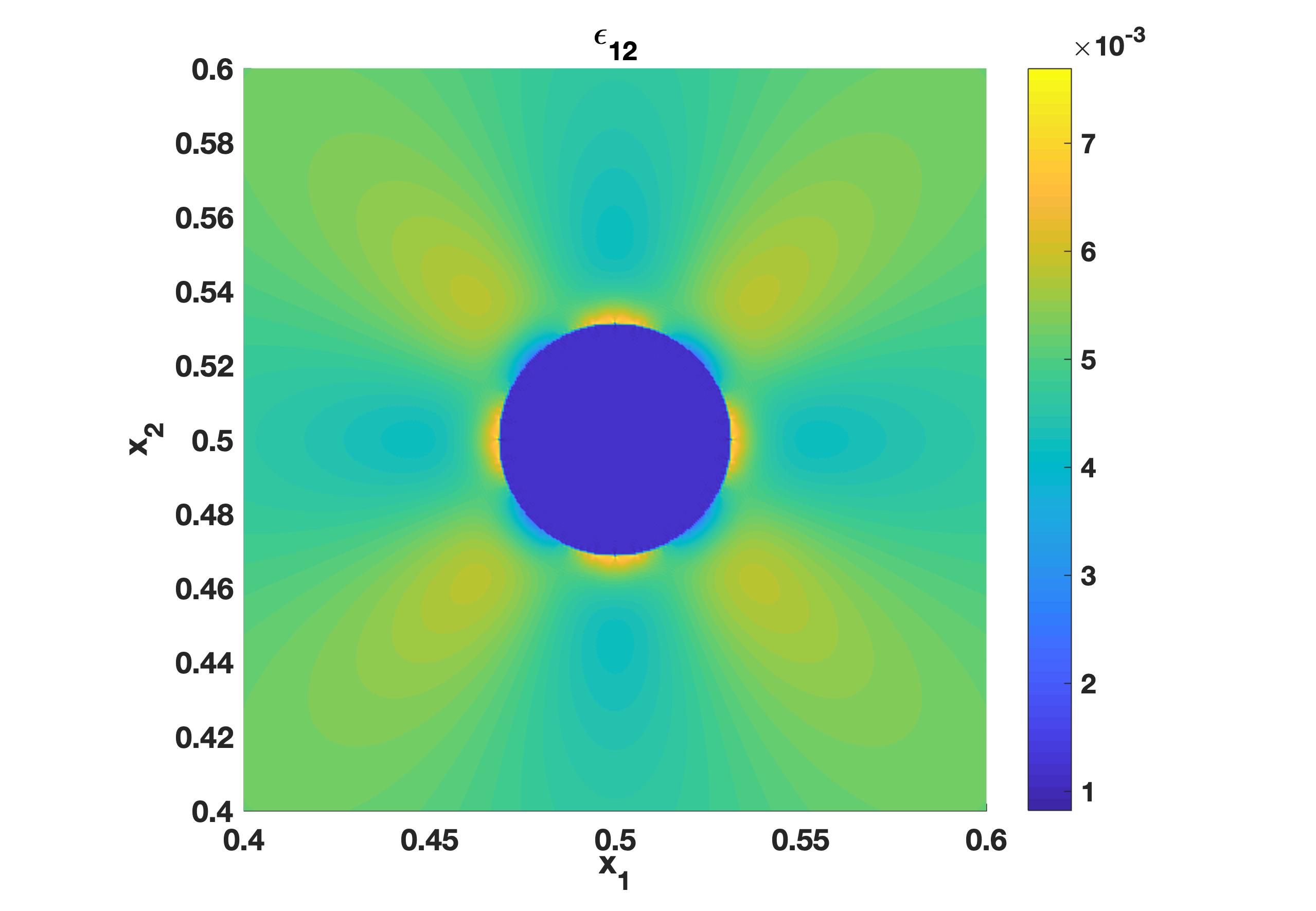}
                    \includegraphics[width=80mm]{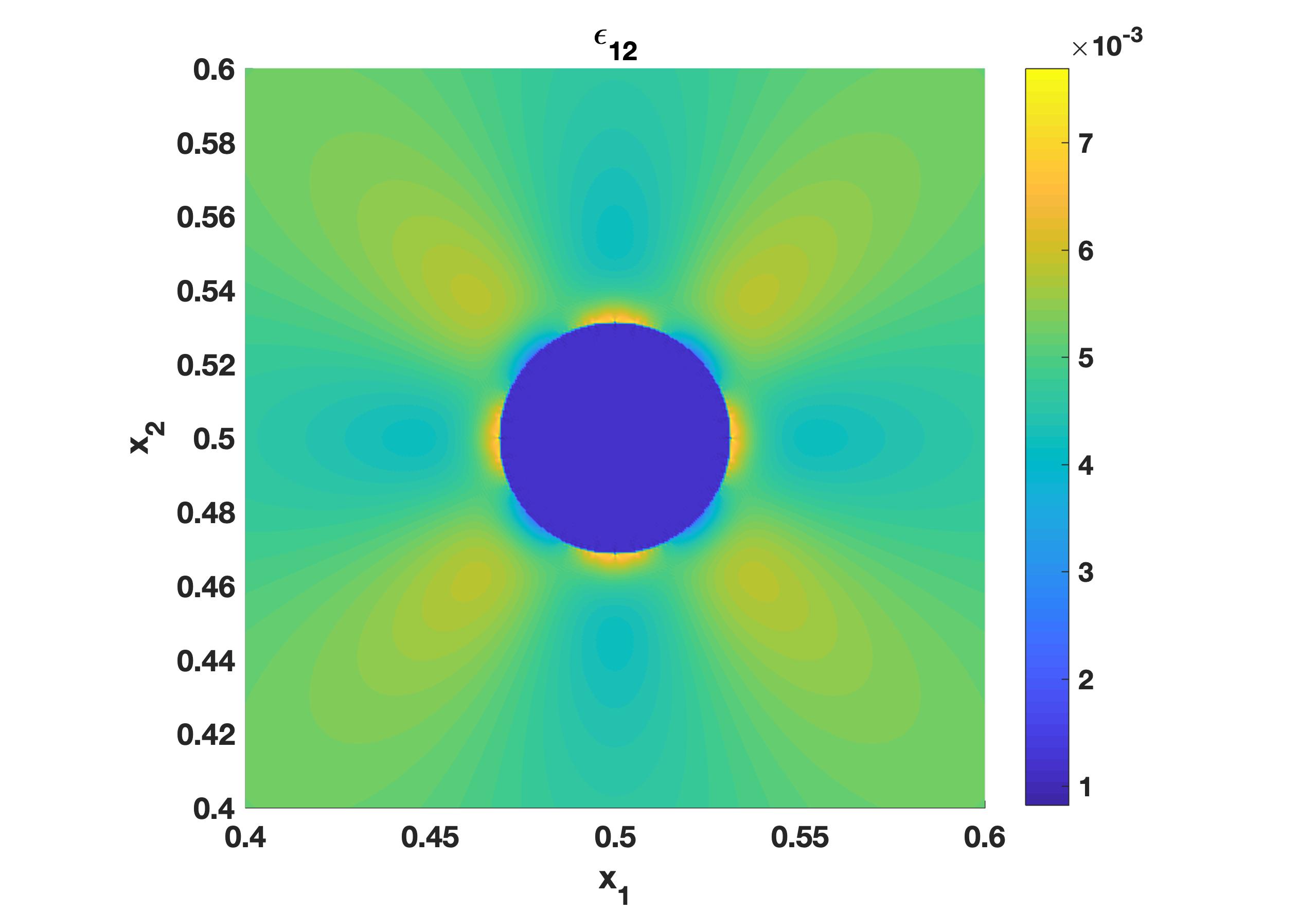}
                    \\
                    \includegraphics[width=80mm]{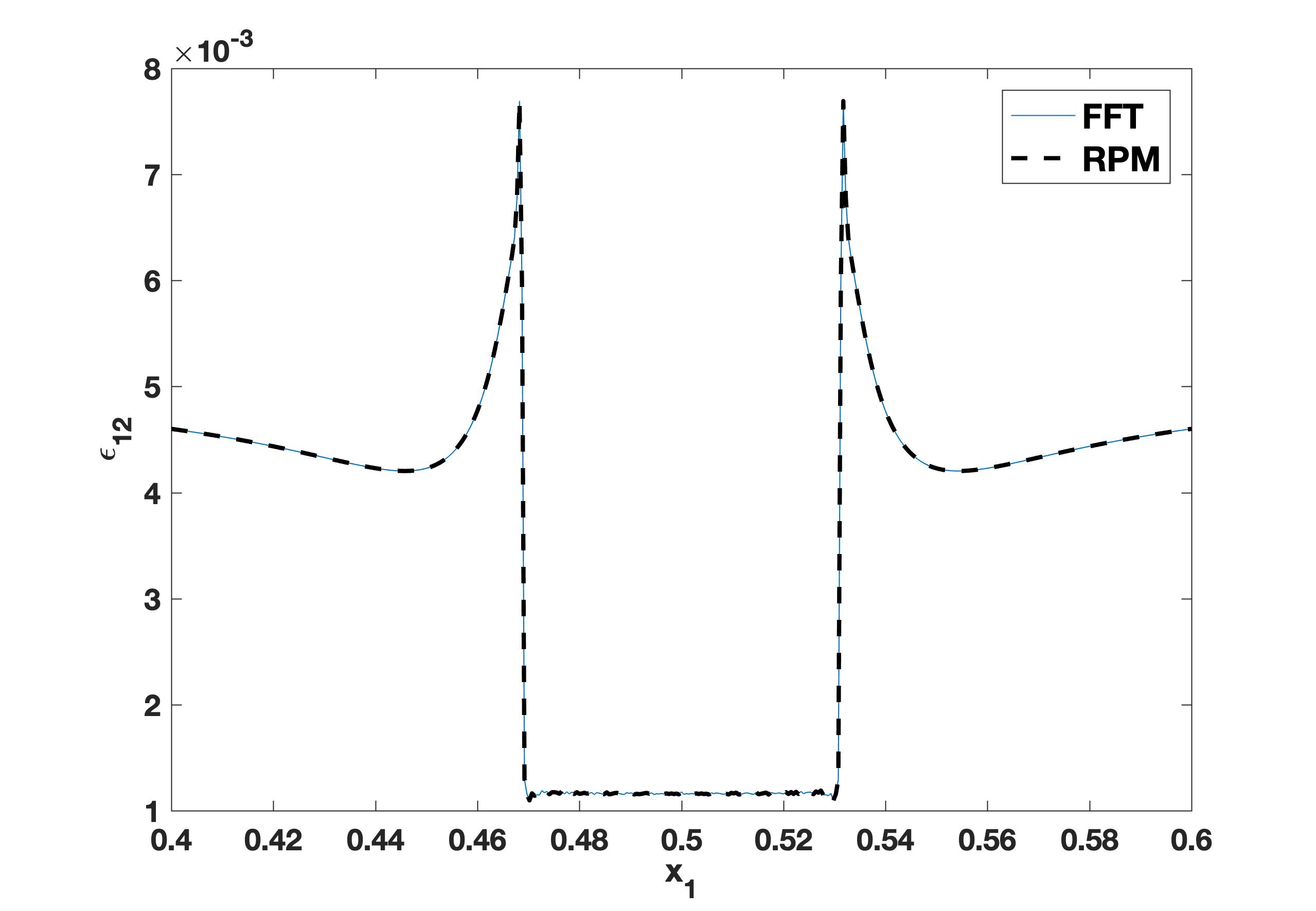}
                    \includegraphics[width=80mm]{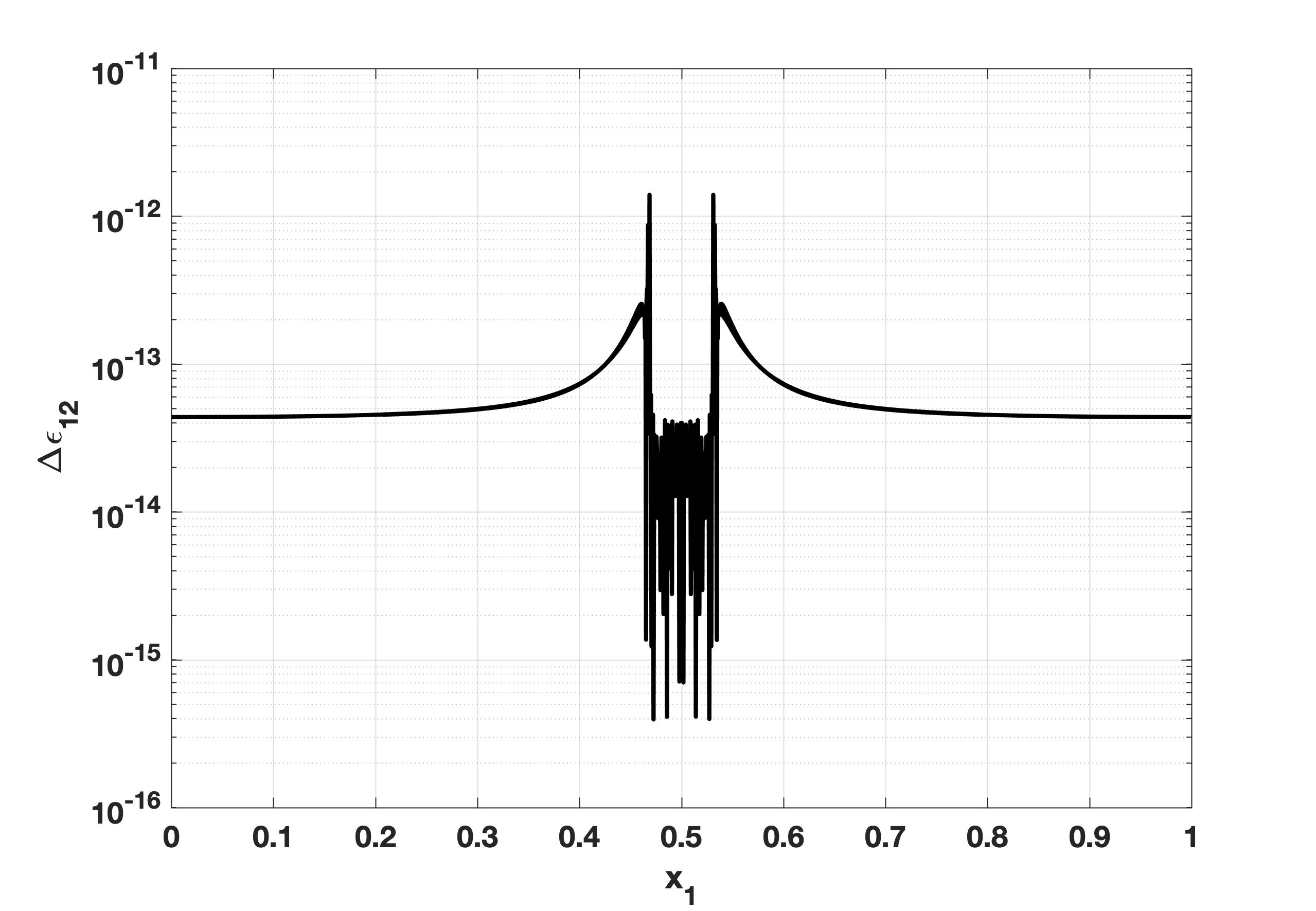}

  					\caption{Top row: $\epsilon_{12}$ component of the strain using Classical FFT (left) and RPM-FFT (center) methods.  Bottom row: the figure on the right shows $\epsilon_{12}$ along the horizontal symmetric axis for both RPM-FFT and Classical FFT; the figure on the left shows the normalized difference. The geometry is given by  $a / b=1/16$, and the figure shows only the region in the vicinity of the fiber.  The mesh is $2048 \times 2048$.  We use  $E_{f}=68.9$GPa and $E_{m}=400$GPa; the contrast $K \approx 5.8$.  We apply shear loading: $E_{12}=0.5\%$, $E_{11}=E_{22}=0$.
						}
					\label{fig:first-test}
				\end{center}
			\end{minipage}
		}
	\end{center}
\end{figure}

As a canonical problem for many of the calculations below, we follow \cite{moulinec1998numerical} in using a circular stiff fiber embedded in a compliant matrix, at the center of a square unit cell and periodic boundary conditions (Fig. \ref{fig:first-test}).
This example is chosen because of the availability of approximate closed-form solutions.
We use the following notation: $a$ denotes the fiber radius; $b$ denotes the size of the square unit cell; $E_f$ and $\nu_f=0.35$ denote respectively the elastic modulus and the Poisson's ratio of the fiber, noting that we assume isotropic elasticity; $E_m$ and $\nu_m=0.23$ denote respectively the elastic modulus and the Poisson's ratio of the matrix, again assuming isotropic elasticity; and $K = E_f / E_m$ is the elastic contrast, where particular focus will be on large values of $K$.
The average strain is denoted by the components $E_{12}, E_{11}, E_{22}$.
For all tests in this section except Section 5.5, the reference medium is assumed to be isotropic and we choose the elastic modulus $\displaystyle E_o =  \frac{E_f + E_m}{2}$ and the Poisson's ratio $\nu_o = \nu_f  = \nu_m$.

As a first test, we simply confirm that RPM-FFT converges to the correct solution in Fig. \ref{fig:first-test}.
There is no discernible difference\footnote{The tolerance in all calculations reported in this paper is taken to be $10^{-4}$.} in the solutions obtained from the Classical FFT and RPM-FFT methods.
Further, Classical FFT takes less time\footnote{All times reported in this paper are wall-clock times.} than RPM-FFT.
This is unsurprising given that the elastic contrast is small ($K \approx 5.8$).
As $K$ increases, the advantages of the RPM-FFT method over the Classical FFT method become clear.
We examine this and other issues below, with a more detailed reporting of computational time.

\subsection{Time to Convergence vs. Elastic Contrast}

We next compare the time taken for convergence across the Classical FFT, Accelerated FFT, and RPM-FFT methods.
We continue with the circular fiber in matrix setting, with $K$ going from $10$ to $10^5$, and $\nu_m = \nu_f = 0.25$.
We use a $256\times 256$ discretization, and apply an average shear strain: $E_{12}=0.5\%$, $E_{11}=E_{22}=0$.

\begin{figure}[h]
    \centering
    \includegraphics[width=100mm]{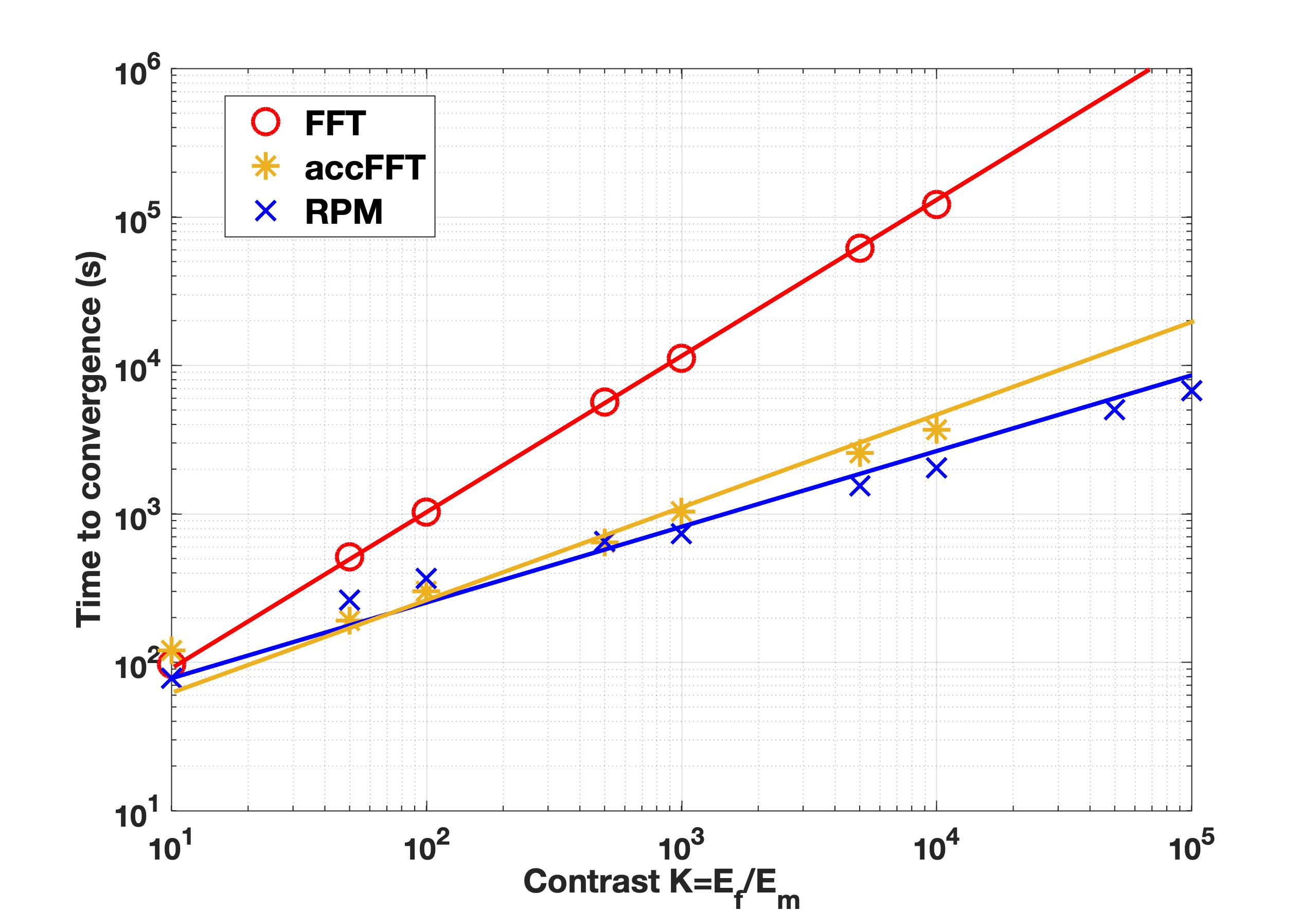}
    \caption{\label{fig:compr3} Time at convergence of three methods with varying $K$. Note the log scale.  The lines are linear best fits.}
\end{figure}

Fig. \ref{fig:compr3} shows the now well-known dramatic improvement between the Classical FFT and the Accelerated FFT methods; \cite{michel2001computational} show that the rate of convergence in the Classical FFT method goes as the contrast $K$, while for the Accelerated FFT method the rate of convergence goes as the square root of the contrast $K$.

Fig. \ref{fig:compr3} also compares the Accelerated FFT method and RPM-FFT method. 
When the contrast is below $10^3$, the Accelerated FFT method converges faster.
However, as the contrast increases, RPM-FFT is increasingly competitive, and surpasses Accelerated FFT.
The rate of convergence of RPM goes as $K^{0.43}$ for this particular class of examples.
The exponent appears to be roughly the same in a few other problems that we tested.

We note that the comparisons for all of these methods was conducted by implementing the codes in Matlab with no additional optimization of any of the methods to ensure fair comparisons.


\subsection{Balancing Newton Iterations against Fixed Point}

As can be expected, the speed of RPM-FFT depends strongly on the size of the unstable subspace on which Newton iterations must be performed.
While the Newton's method has better convergence properties, each Newton iteration is also far more expensive than the fixed point iteration, particularly as the dimension of the unstable subspace increases.
On the other hand, the fixed point method converges slowly, or not at all, on the unstable subspace, which also increases the expense as it requires a large number of iterations.

In this section, the term ``unstable subspace'' refers to the subspace that the RPM-FFT method identifies; this is not necessarily the true unstable subspace whose corresponding eigenvalues are outside the unit disk.

An optimal method has to balance between the opposing issues identified above.
Denoting by $n_{max}$ the number of fixed-point iterations before increasing the size of the unstable subspace, we notice that setting $n_{max}$ too low causes the dimensionality of the unstable subspace to increase quickly and makes each Newton iteration expensive; on the other hand, if $n_{max}$ is too high, there will be minimal progress from the large number of fixed-point iterations before switching to Newton iterations.

\begin{figure}[htb]
    \centering
    \includegraphics[width=0.8\textwidth]{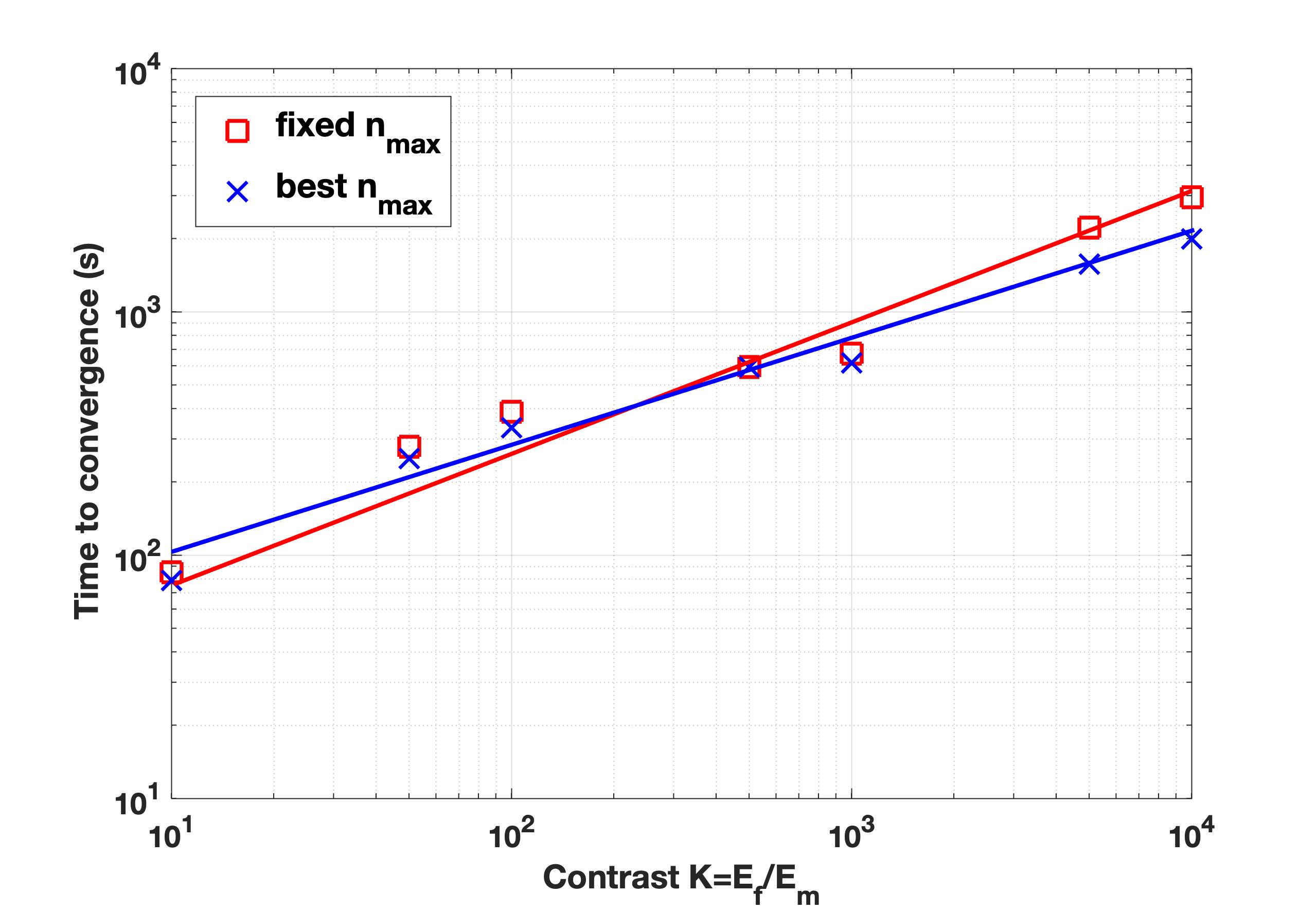}
    \caption{\label{fig:fx/bs} Time at convergence of RPM with fixed $n_{max}=10$ and $n_{max}^{best}(K)$.  The lines are linear best fits.}
\end{figure}

We examine this interplay numerically for the model problem described previously, for varying $K$.
We first fix $K$, and then compare the times obtained for a large range of $n_{max}$ going from $1$ to $50$.
We then repeat over a range of $K$.
Of these, we denote the value of $n_{max}$ with the best (least) time as $n_{max}^{best}(K)$.
Fig. \ref{fig:fx/bs} shows the times to convergence for a fixed value of $n_{max} = 10$, as well as the times to convergence for $n_{max}^{best}(K)$, as a function of $K$.
An immediate conclusion is that there is almost no difference at low $K$, and not much difference at higher $K$.

\begin{figure}[htb]
    \centering
    \includegraphics[width=0.65\textwidth]{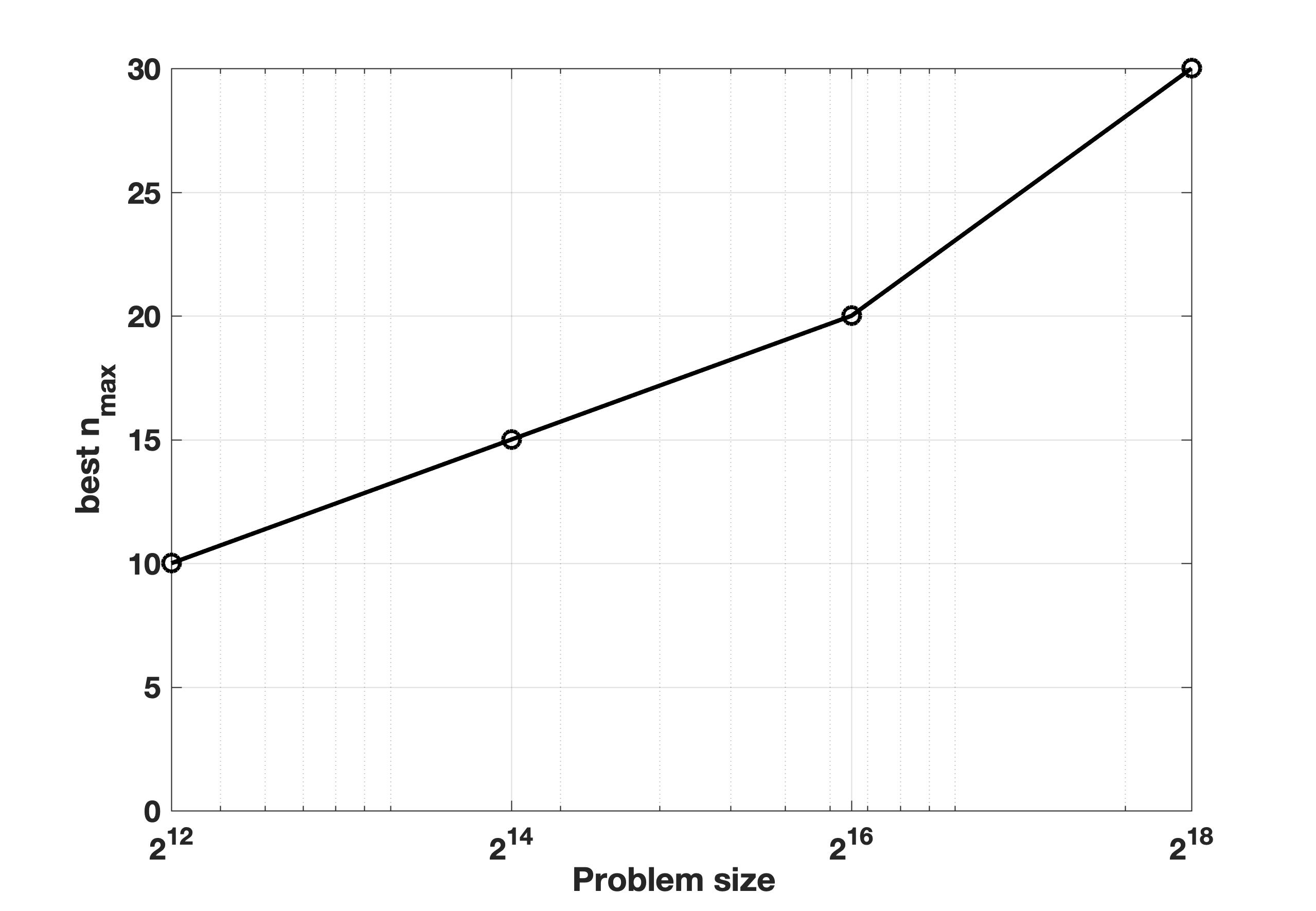}
    \caption{\label{fig:res/bnx} The best $n_{max}$ for different problem sizes. The line connects datapoints to aid visualization.}
\end{figure}

We next compare the effect of the size of the problem on $n_{max}^{best}(K)$, for a fixed value of $K=50$.
Fig. \ref{fig:res/bnx} shows $n_{max}^{best}$ for different problem sizes, here identified with the number of grid points in each direction.
An immediate conclusion is that $n_{max}^{best}$ increases with the problem size.

\subsection{Effect of Volume Fraction and Problem Size}

In the model problem of a stiff fiber in a soft matrix, the volume fraction of the stiff fiber has been taken to be small; for $a/b = 1/16$, the stiff fiber has a volume fraction $\lambda = 0.3\%$.  Note that we use the term volume fraction though we are working in the 2D setting.

We expect that the size of the unstable subspace is related to the size of the fiber based on the heuristics discussed in Sections \ref{sec:intro}, \ref{sec:RPM-FFT} and \ref{sec:variation}.
As the volume fraction increases, we therefore expect a a larger unstable subspace and consequently more expensive Newton iterations.

\begin{figure}[h]
    \centering
    \includegraphics[width=0.8\textwidth]{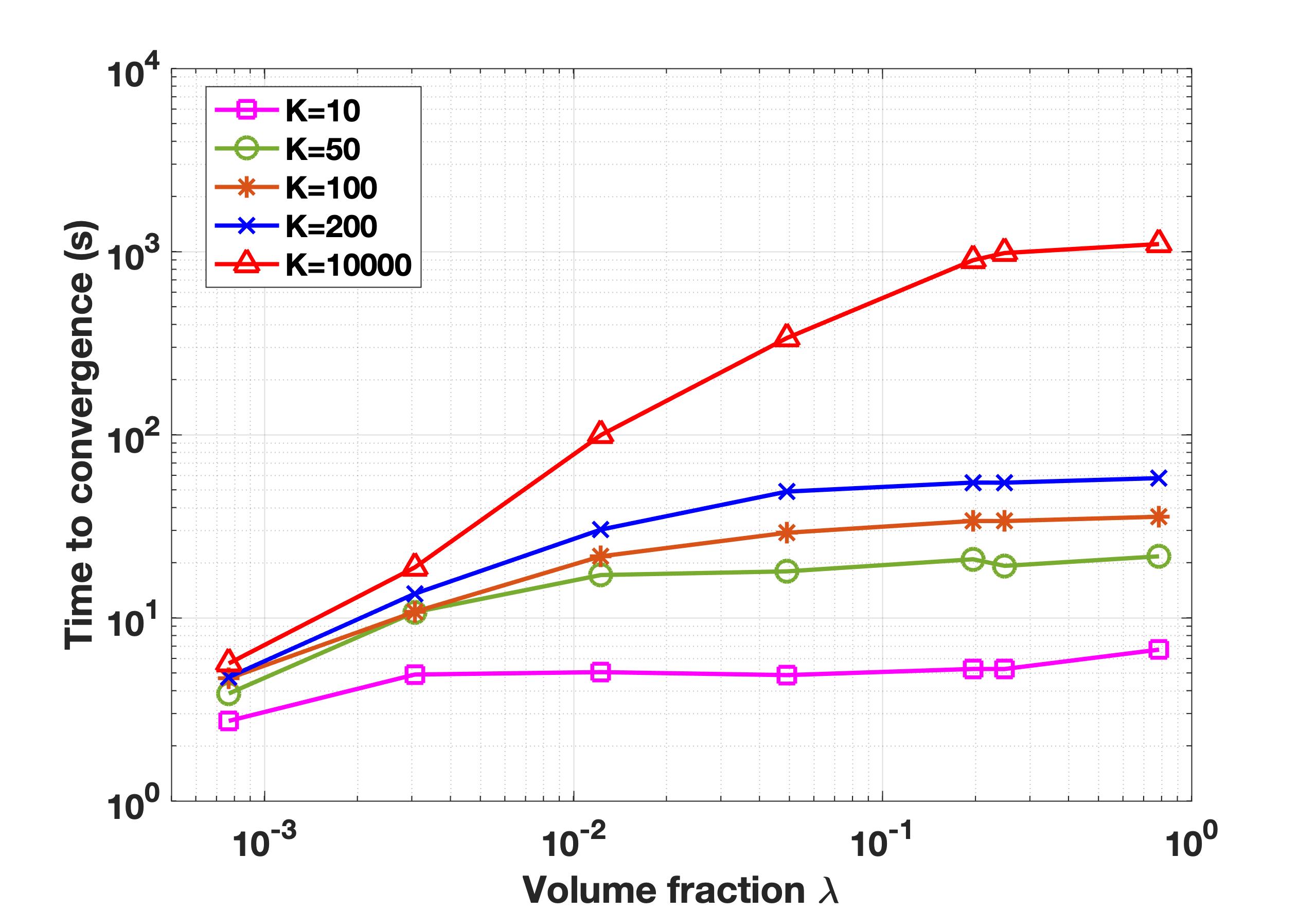}
    \caption{\label{fig:gm/K_T} Time to convergence of RPM-FFT with varying $\lambda$ and $K$.  The lines connect the datapoints to aid visualization.}
\end{figure}

We examine this issue through numerical experiments, and find that RPM-FFT works well even at the combination of large elastic contrast $K$ and volume fraction of fiber $\lambda$.
For instance, even with $K\approx 10^4$ or higher, and $a/b=9/10 \Rightarrow \lambda = 63.6\%$, RPM-FFT is easily able to converge.
Fig. \ref{fig:gm/K_T} shows the time to convergence as a function of $\lambda$.
While larger elastic contrasts require more time to converge -- in line with our observations above -- we notice that the time to converge is almost insensitive to $\lambda$ for volume fractions larger than $5\%$.

\begin{figure}[h]
    \centering
    \includegraphics[width=0.8\textwidth]{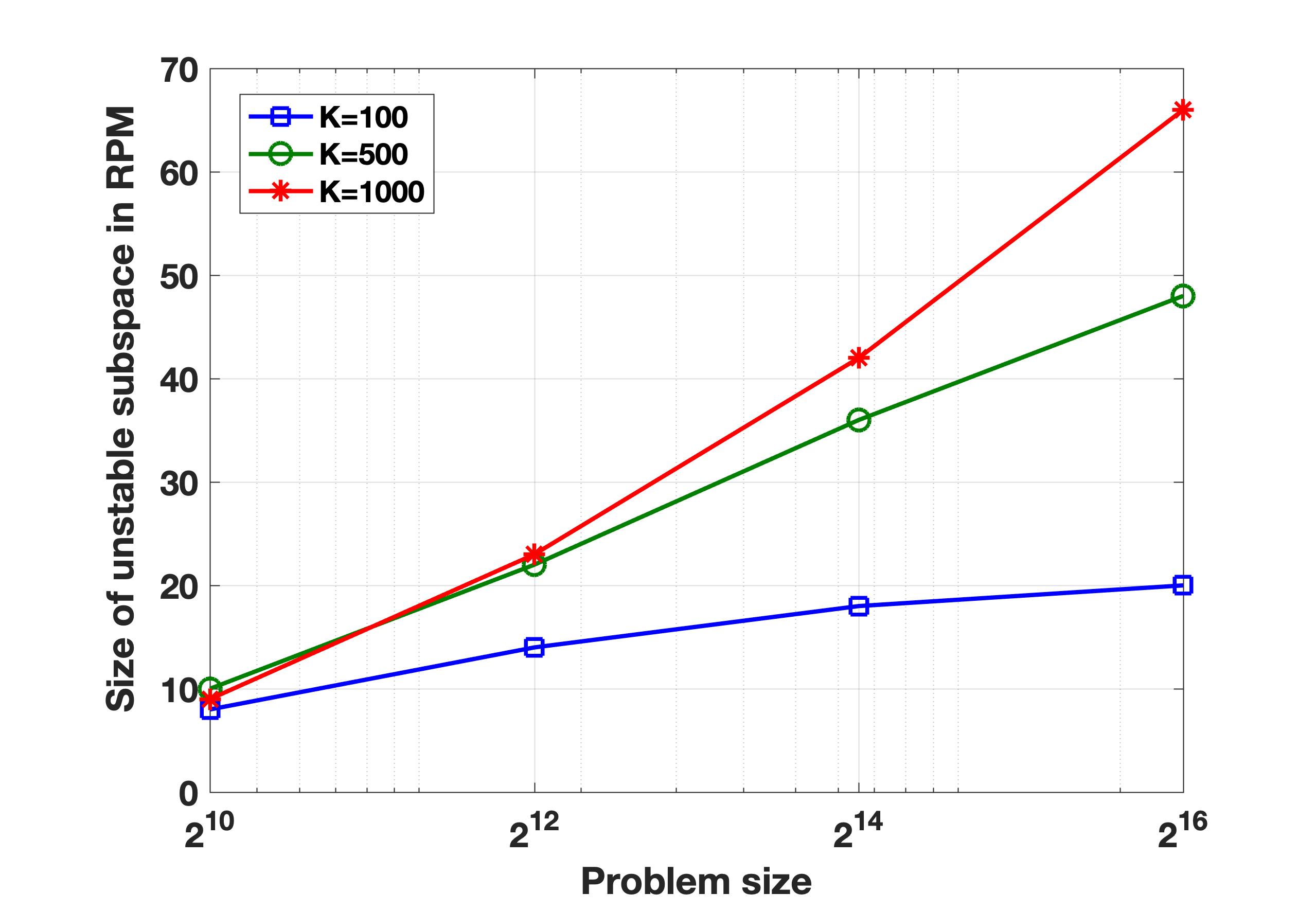}
    \caption{\label{fig:uns_pt} Size of unstable subspace with problem size.  The lines connect the datapoints to aid visualization.}
\end{figure}

We next consider the case of fixed $\lambda$ but with changing problem size with a fixed geometry.
That is, we compare discretizations of varying coarseness.
The number of grid points within the fiber, as well as overall, increases as the discretization is made finer.
We consequently expect the unstable subspace identified by RPM to also increase.
Fig. \ref{fig:uns_pt} shows the relation between the size of the unstable subspace as the problem size increases from $32\times 32$ to $512 \times 512$, for different values of elastic contrast $K$.
The volume fraction $\lambda$ is held fixed $1.2\%$.
Roughly, we observe a logarithmic scaling between the size of the unstable subspace and the number of grid points within the fiber.

\subsection{Interaction Between Stiff Fibers}

Motivated by recent advances in functional composites \cite{Dhriti,roy2017review}, we examine the setting of two circular carbon fibers embedded in a polyurethane matrix.
It is a significant challenge to characterize the effective properties of heterogeneous materials such as nanofiber- and nanoparticle- reinforced nanocomposites starting from the scale of resolving the entire microstructure.
This makes it challenging to develop physics-based high-fidelity model that predict the performance of such nanocomposite systems. 
The FFT methods can help to make stronger direct connections between mechanics modeling and experimental data, which in turn can have a significant impact on understanding the structure-property relations of these complex heterogeneous nanocomposite systems.

\begin{figure}[h]
	\begin{center}
		\fbox{
			\begin{minipage}{175mm}
				\begin{center}
                    \includegraphics[width=85mm]{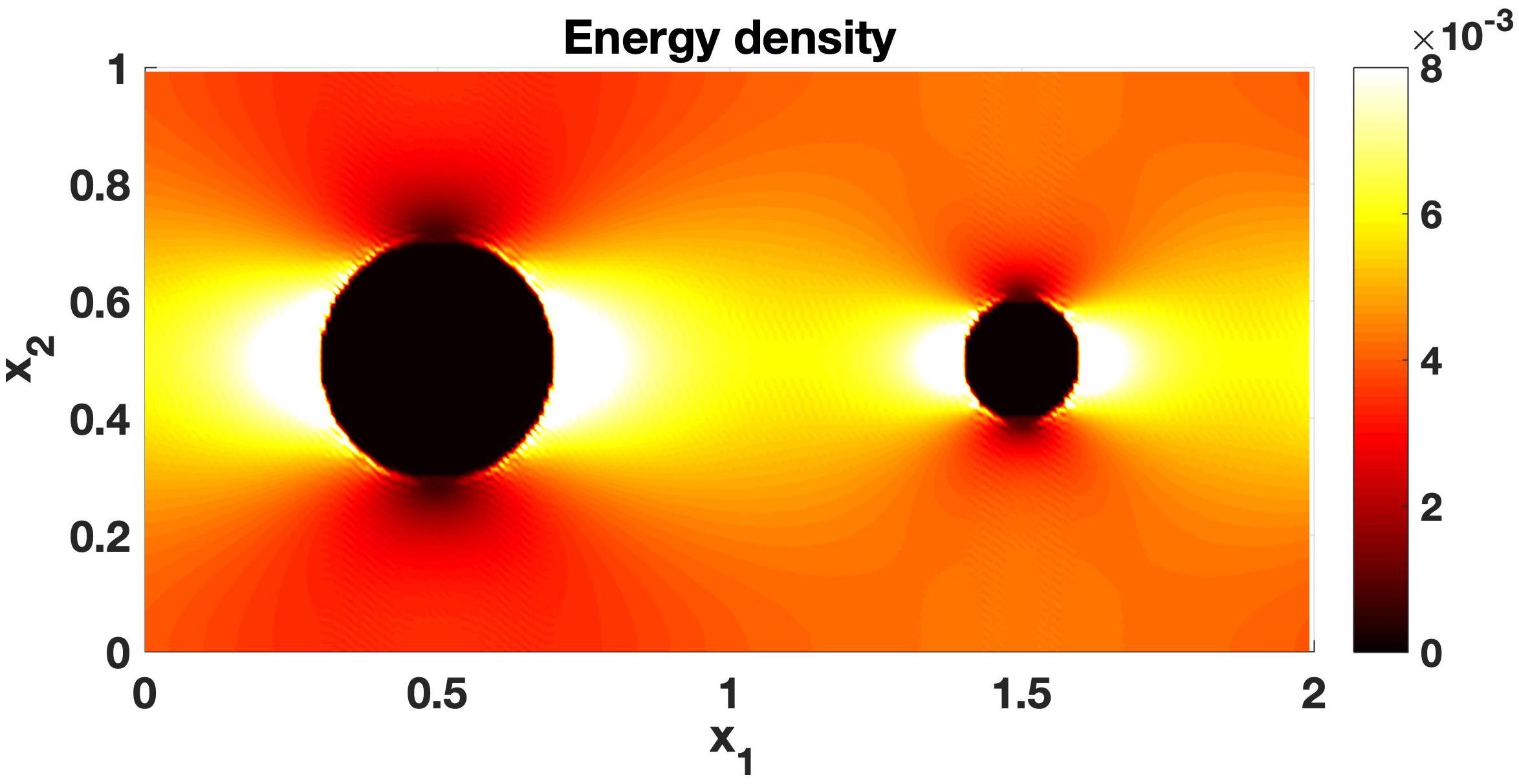}
                    \includegraphics[width=85mm]{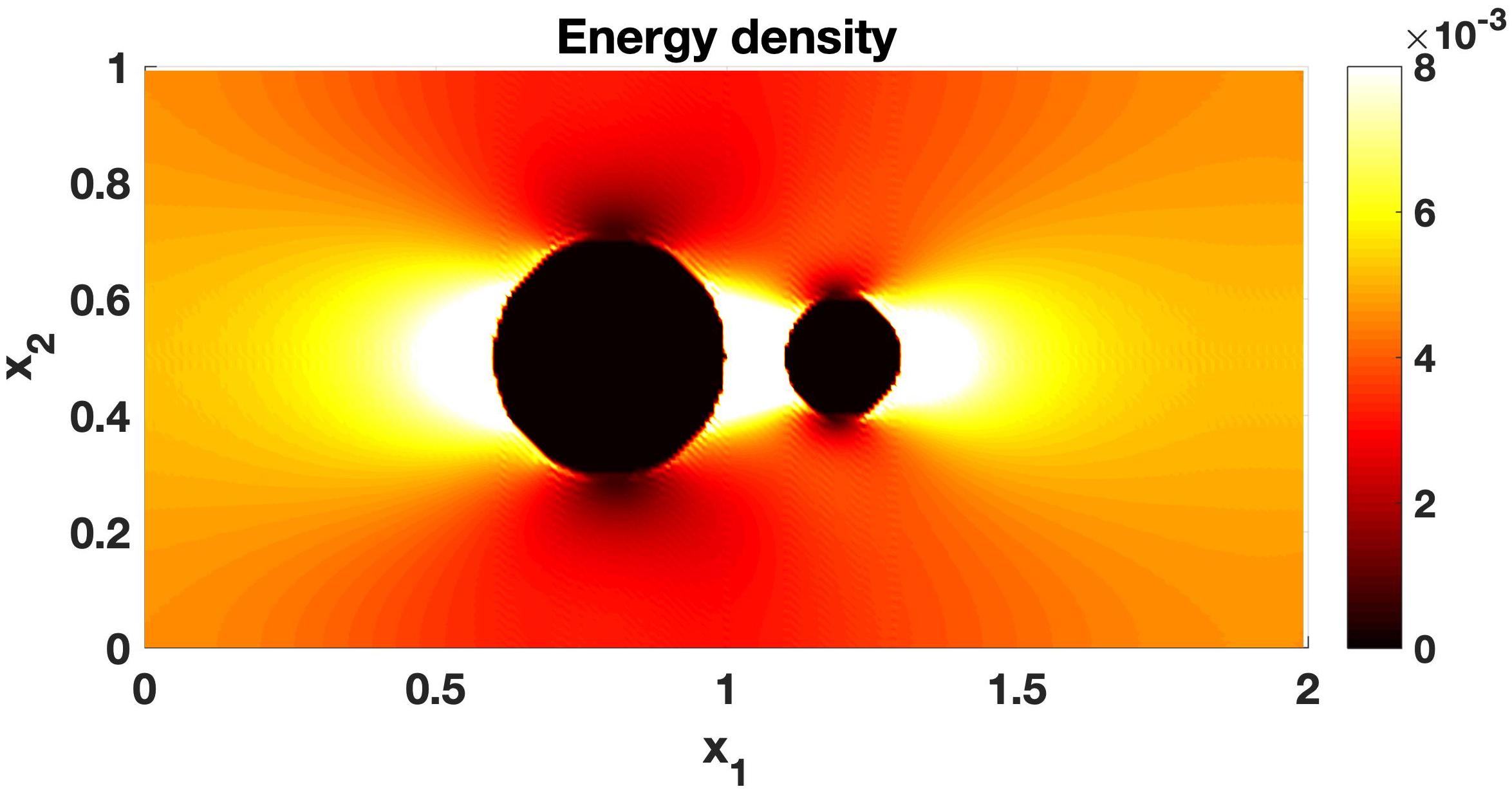}
					\\
                    \includegraphics[width=85mm]{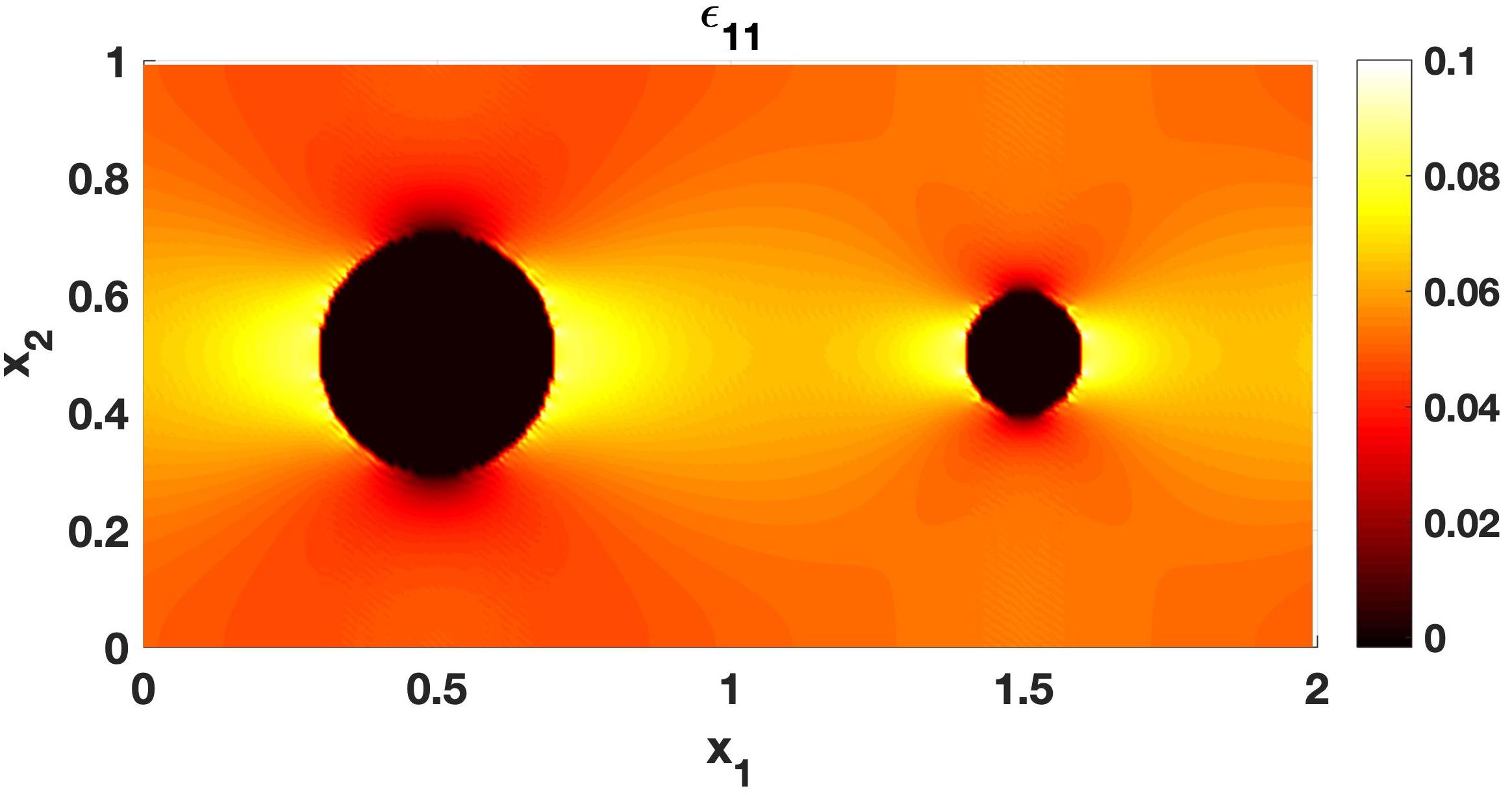}
                    \includegraphics[width=85mm]{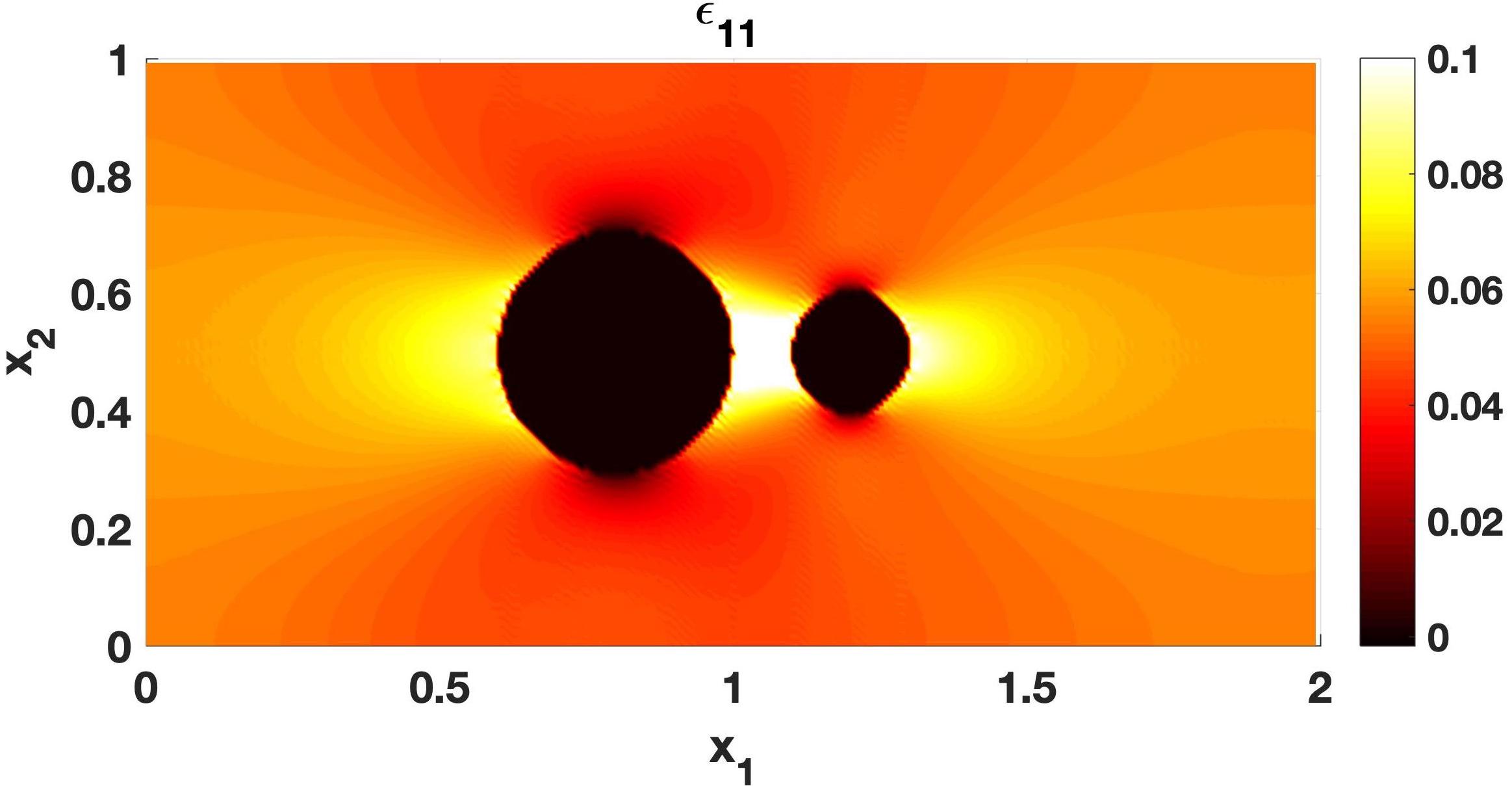}
					\caption{Energy density (top row) and $\epsilon_{11}$ (bottom row) for almost-rigid fibers, when the fibers are near and far away from each other.
						}
					\label{fig:2-fibers}
				\end{center}
			\end{minipage}
		}
	\end{center}
\end{figure}

We consider the geometry of Fig. \ref{fig:2-fibers} with two fibers that are either close to each other -- the separation is smaller than the smaller fiber diameter -- and when they are further apart.
The moduli are assumed to be $E_f = 900$ GPa for the carbon fiber and $E_m = 0.03$ GPa for the polyurethane, giving an elastic contrast of $K=3\times10^4$.
Two fibers, radius $a$ and $2a$, are located in a rectangular unit cell of size $2b \times b$, with $b=10a$.
We examine two cases: when the fibers are far apart with distance $b$ between the centers of the fibers, and when the fibers are closer with distance $0.4 b$, see Fig. \ref{fig:2-fibers}.
 We apply a uniaxial average strain: $E_{11}=5\%, E_{22}=E_{12}=0$.


Fig. \ref{fig:2-fibers} shows strain and strain energy density fields for the two configurations.
In all of these configurations, the strain and strain energy density fields are essentially $0$ within the stiff fibers, and also show good convergence properties with no obvious spurious oscillations near the transition from matrix to fiber.
We notice that in the case where the fibers are further apart, the energy density and strain fields suggest that the fibers can be essentially considered dilute and non-interacting.
However, when the fibers are closer together -- as is typical of experimental specimens \cite{Dhriti,roy2017review} -- there is a clear interaction between the fields, and the energy density and strain fields take their largest values in the region between the fibers.


\subsection{RPM Applied to the Accelerated FFT Method}
Two important advantages of the RPM approach are, first, that it uses the fixed-point iterations to identify the unstable subspace; and, second, once identified, it performs Newton iterations on the unstable subspace with fixed-point iterations on the {\em entire} space.
This makes it easy to apply the RPM strategy to existing fixed-point methods with minimal changes to algorithm / code.
We therefore exploit this to apply the RPM approach to the Accelerated FFT method which is also based on a fixed-point framework.
Given that the Accelerated FFT method has much superior performance compared to the Classical FFT method, we might expect that a RPM-Accelerated FFT method likely performs even better.
We examine this issue below.

In \cite{moulinec2014comparison}, it is shown  that the Accelerated FFT method is a special case of the general polarization-based fixed-point method developed by  \cite{monchiet2012polarization}. 
Algorithm \ref{alg:pol} is the general polarization-based method, where $\alpha$ and $\beta$ can take any value within a range.
For the choice $\alpha=\beta=2$, the polarization-based method reduces to the Accelerated FFT method.  
In \cite{moulinec2014comparison}, they also discuss and compare the Accelerated FFT method with other polarization-based methods, corresponding to the choices $\alpha=\beta=1.5$ and $\alpha=\beta=1$. 

All of these methods are fixed-point based methods, and therefore require that the eigenvalues of the Jacobian are within the unit disk for convergence.
The Accelerated FFT method has the lowest upper bound \cite{moulinec2014comparison}, and we therefore combine this with the RPM method.

In all of these methods, an important part of the method is the additive decomposition of the heterogeneous stiffness into a homogeneous reference medium and a fluctuation; see Section \ref{sec:variation}.
The convergence rate is typically very sensitive to the choice of reference medium.

It has been observed that the choice of the reference medium affects the rate of convergence in many fixed-point based schemes. 
In the classical FFT method \cite{moulinec1998numerical}, the best choice was found to be the average of the supremum and infimum of the elastic moduli. 
In Accelerated FFT \cite{michel2001computational} and Polarization FFT \cite{monchiet2012polarization} the optimum is shown to be $E_o / E_f  = \sqrt K$ when computing with a very high precision ($10^{-10}$).

However, as also discussed in Section \ref{sec:variation}, we expect the RPM-based approach to be insensitive to this choice.
Therefore, we apply the RPM approach to the Accelerated FFT method, which is readily obtained by simply replacing the fixed-point operation in algorithm \ref{alg:RPM} with the iteration procedure in algorithm \ref{alg:pol}.

\begin{algorithm}
\caption{Polarization-FFT algorithm}\label{alg:pol}
\begin{algorithmic}[1]

        \State \textbf{Iteration $i+1$} (given $\bfepsilon^i$)
        \State $\bfsigma^{i} \gets \bfC(\bfx):\bfepsilon^{i}$
        \State $\bfs_a^{i}=\bfsigma^i+(1-\beta)\bfC^0:\bfepsilon^i$
        \State $\bfs_b^{i}=\alpha\bfsigma^i+\beta\bfC^0:\bfepsilon^i$
        \State $\hat{\bfs_b}^i \gets \mathcal{F}[\bfs_b^i];$
        
        \State $\hat{\bfepsilon_b}^{i} \gets -\hat{\bfGamma}^0:\hat{\bfs_b}^i \quad\quad$ $\forall \bfk \neq \mathbf{0}$ and $\hat{\bfepsilon_b}(\mathbf{0})^{i}=\beta\bfE$ 
        \State $\bfepsilon_b^{i} \gets \mathcal{F}^{-1}[\hat{\bfepsilon_b}^{i}]$
        \State $\bfepsilon^{i+1}=(\bfC(\bfx)+\bfC^0)^{-1}:(\bfs_a^i+\bfC^0:\bfepsilon_b^i)$
\end{algorithmic}
\end{algorithm}

Table \ref{tab:fourcomp} compares the number of iterations and the time for Accelerated FFT and RPM-Accelerated FFT for a wide range of contrasts.
The enhancement provided by RPM is increasingly large as the elastic contrast $K$ increases.


\begin{table}[]
    \centering
    \begin{tabular}{|r|r|r|r|r|}
        \hline
        \ & \multicolumn{2}{c|}{Number of Iterations} & \multicolumn{2}{c|}{Time (seconds)} \\
        \hline
        K & Acc-FFT & RPM-Acc-FFT & Acc-FFT & RPM-Acc-FFT \\
        \hline
        500 & 234 & 83 & 6.46 & 4.47 \\
        1000 & 335 & 81 & 8.51 & 4.42 \\
        5000 & 335 & 81 & 20.24 & 4.40 \\
        10000 & 1136 & 82 & 28.29 & 4.44 \\
        50000 & 2741 & 82 & 68.56 & 4.45 \\
        100000 & 3863 & 161 & 98.60 & 11.13 \\
        \hline
    \end{tabular}
    \caption{Number of iterations and time to convergence for the Accelerated-FFT and RPM-Accelerated-FFT for a wide range of $K$.}
    \label{tab:fourcomp}
\end{table}

\begin{figure}[ht!]
    \vspace{-10mm}
	\begin{center}
		\fbox{
			\begin{minipage}{175mm}
				\begin{center}
                    \includegraphics[width=80mm]{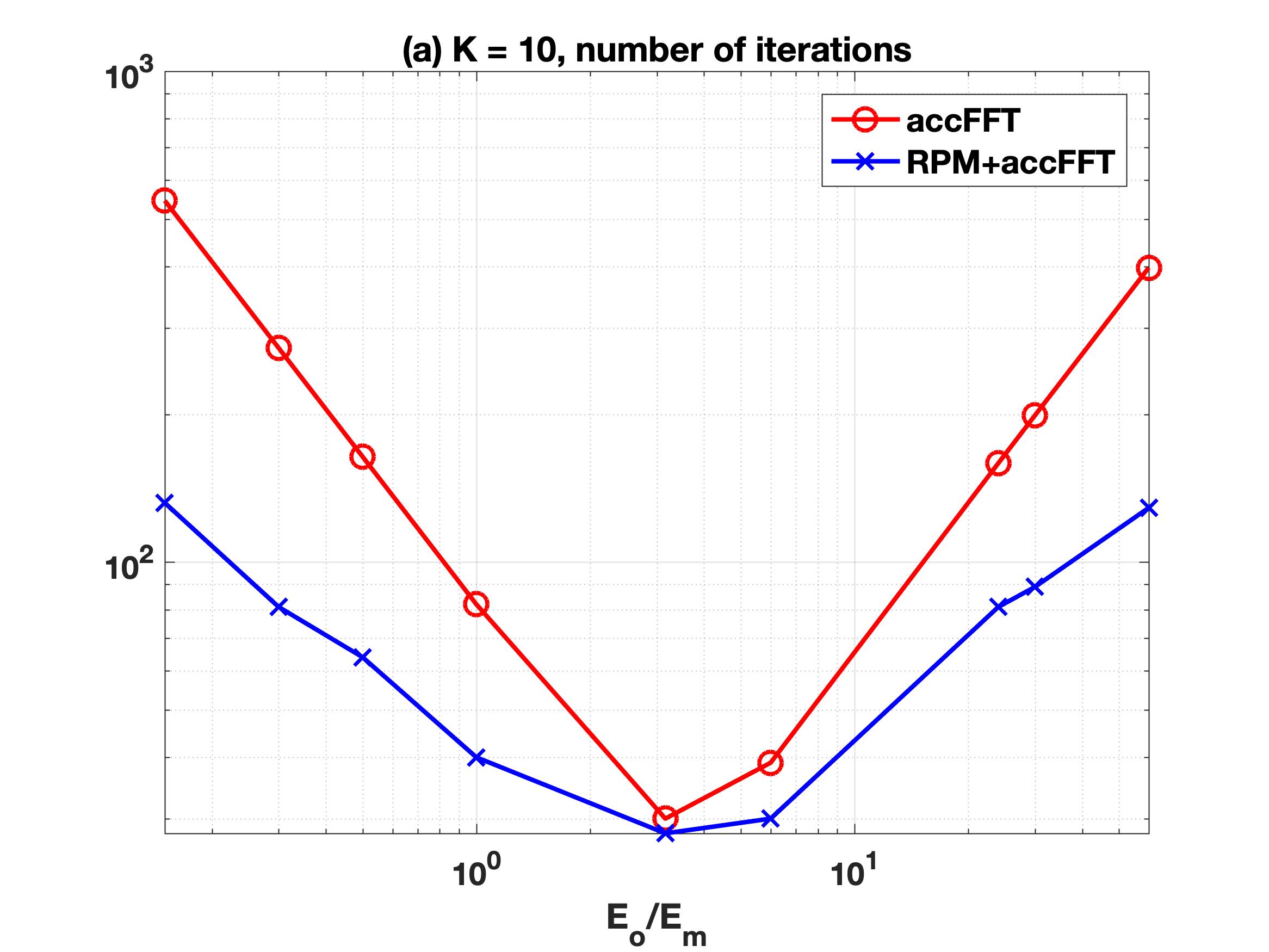}
                    \includegraphics[width=80mm]{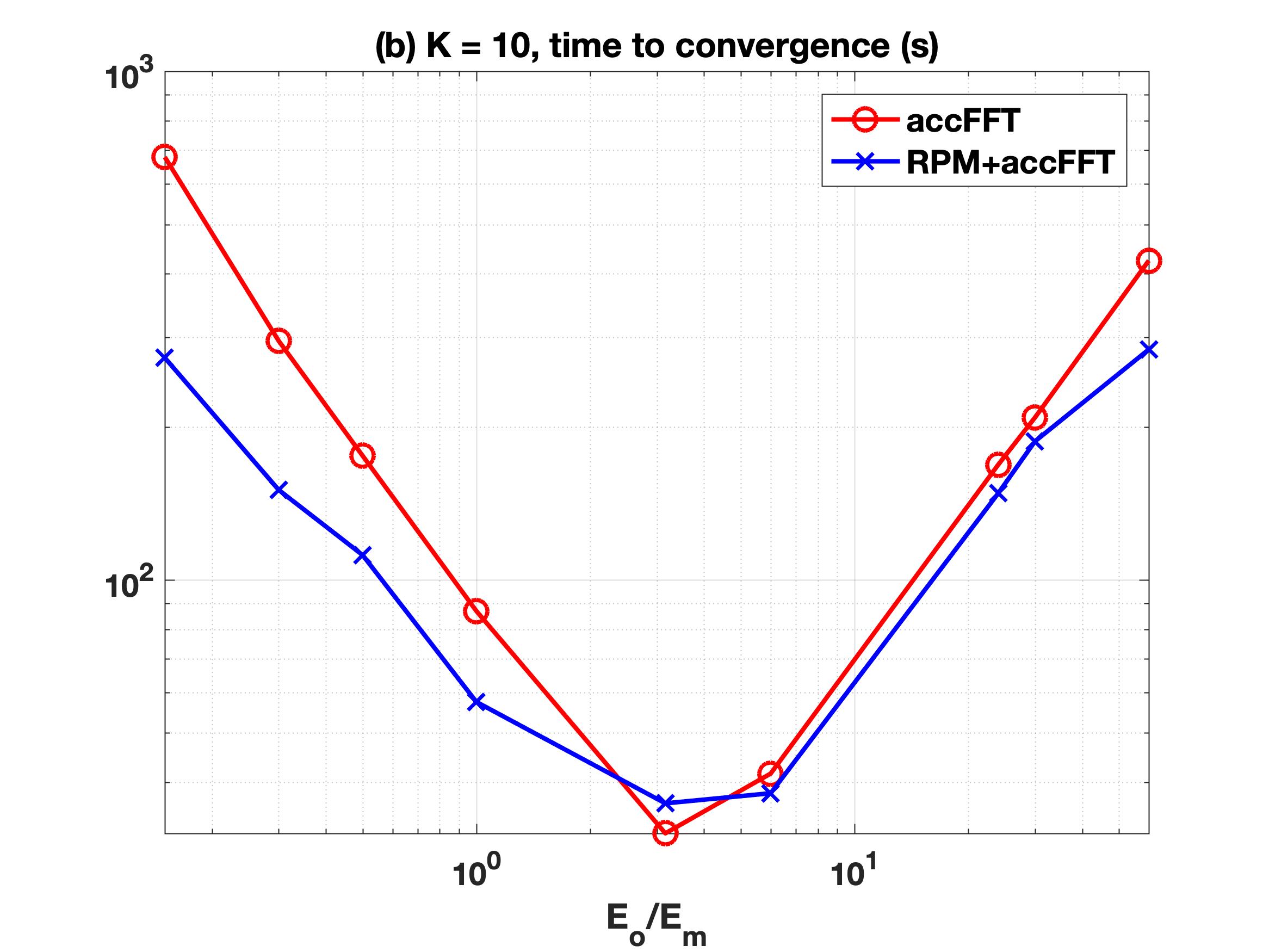}
                    \\
                    \includegraphics[width=80mm]{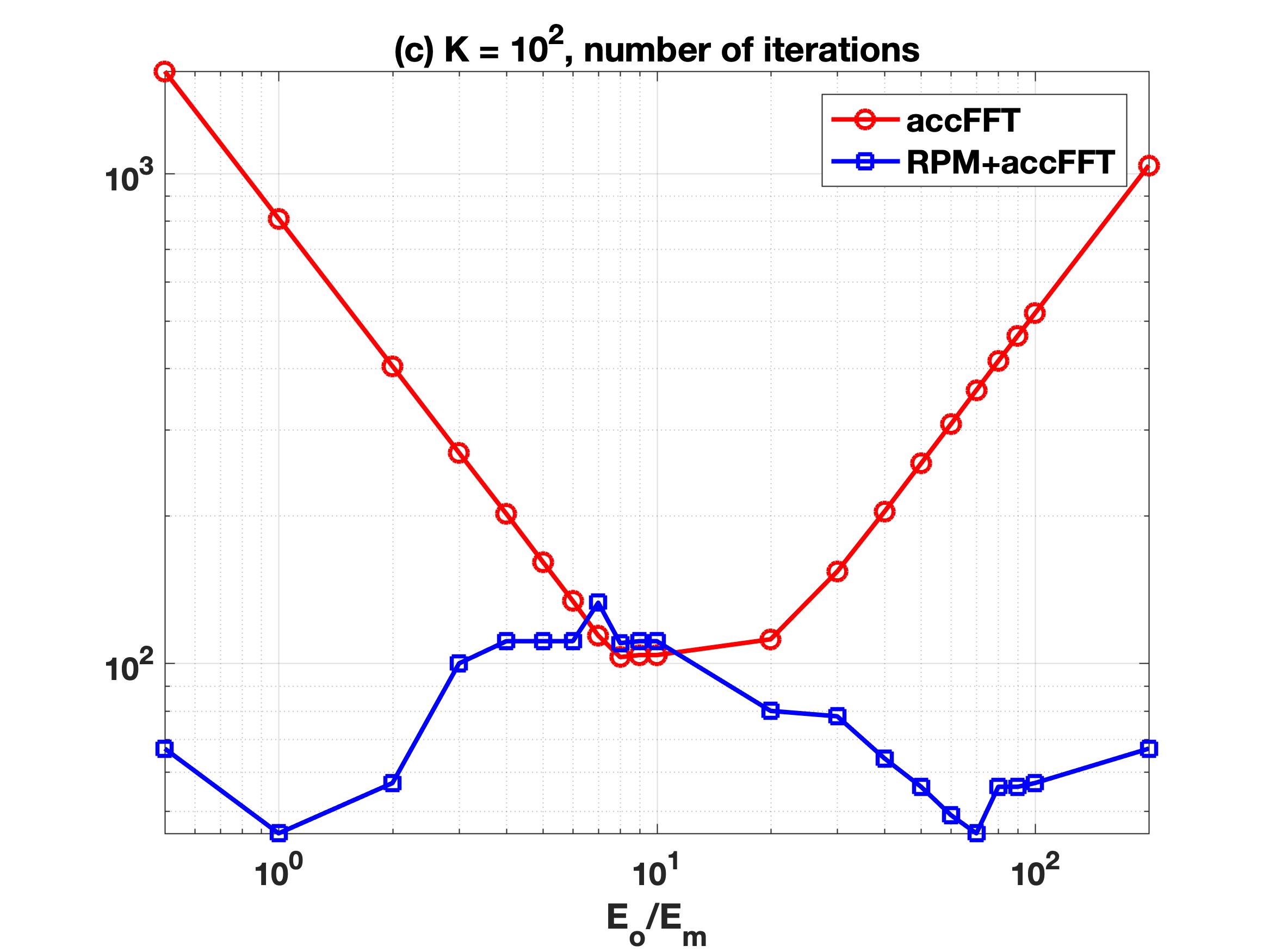}
                    \includegraphics[width=80mm]{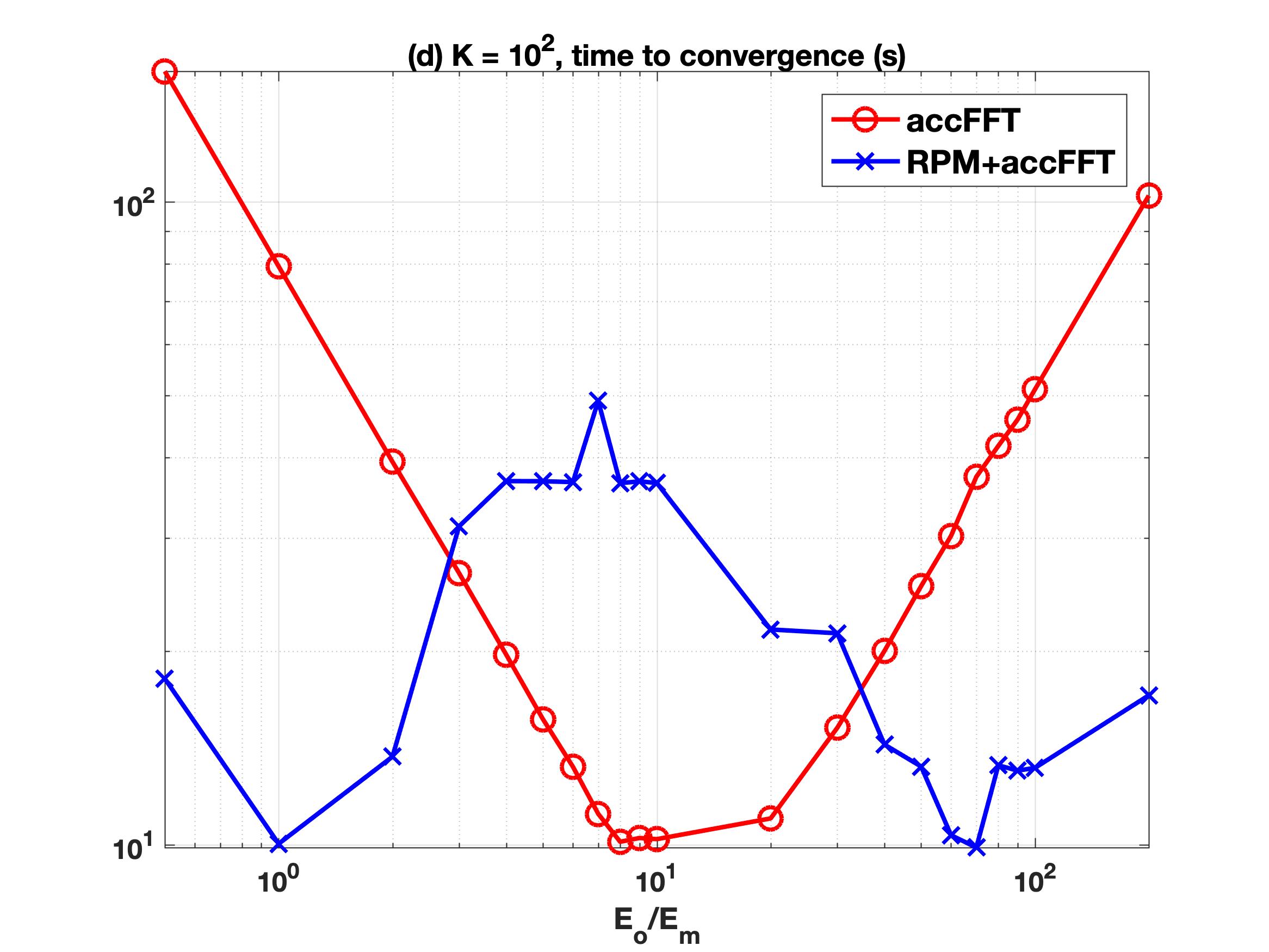}
                    \\
                    \includegraphics[width=80mm]{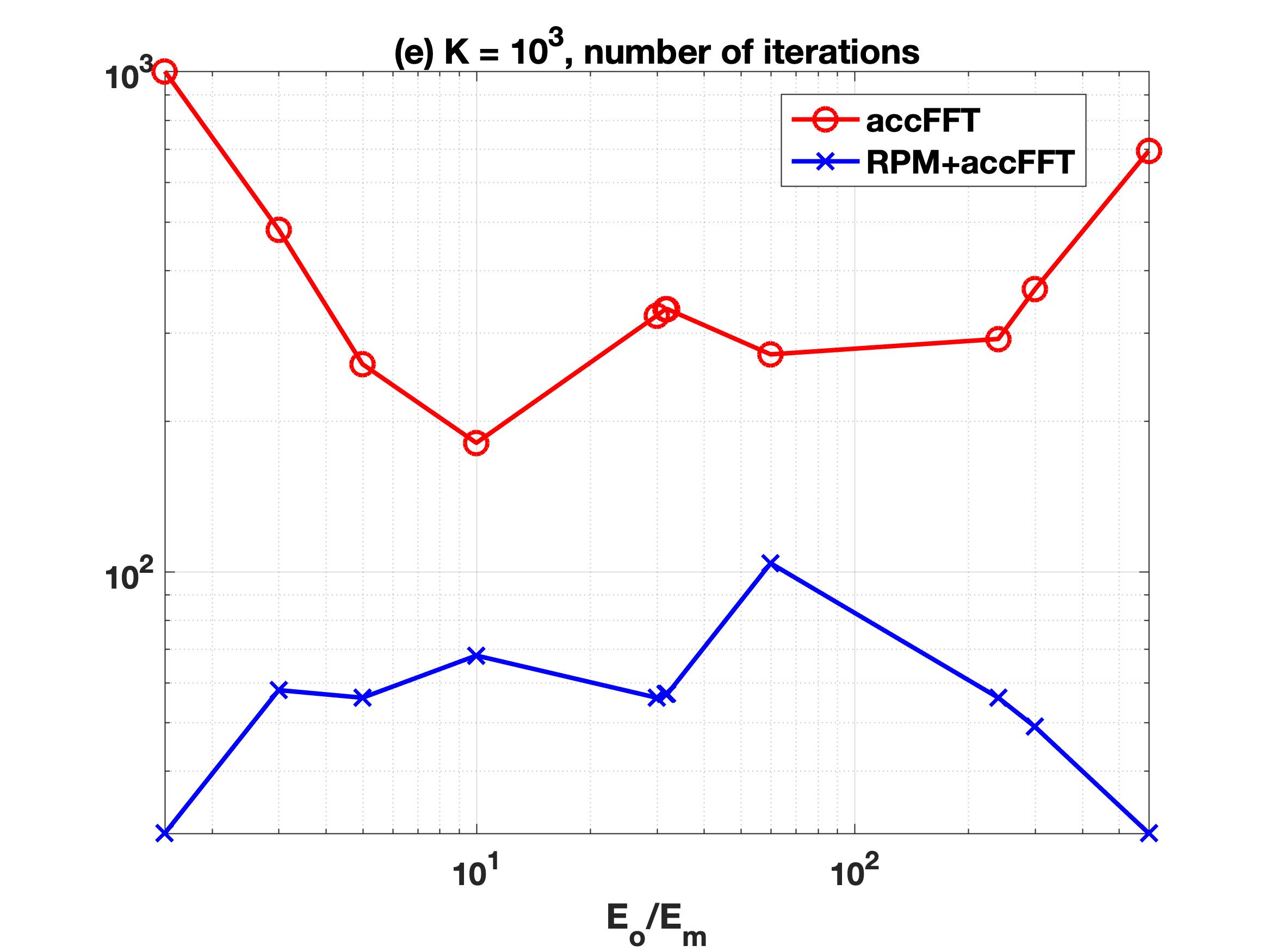}
                    \includegraphics[width=80mm]{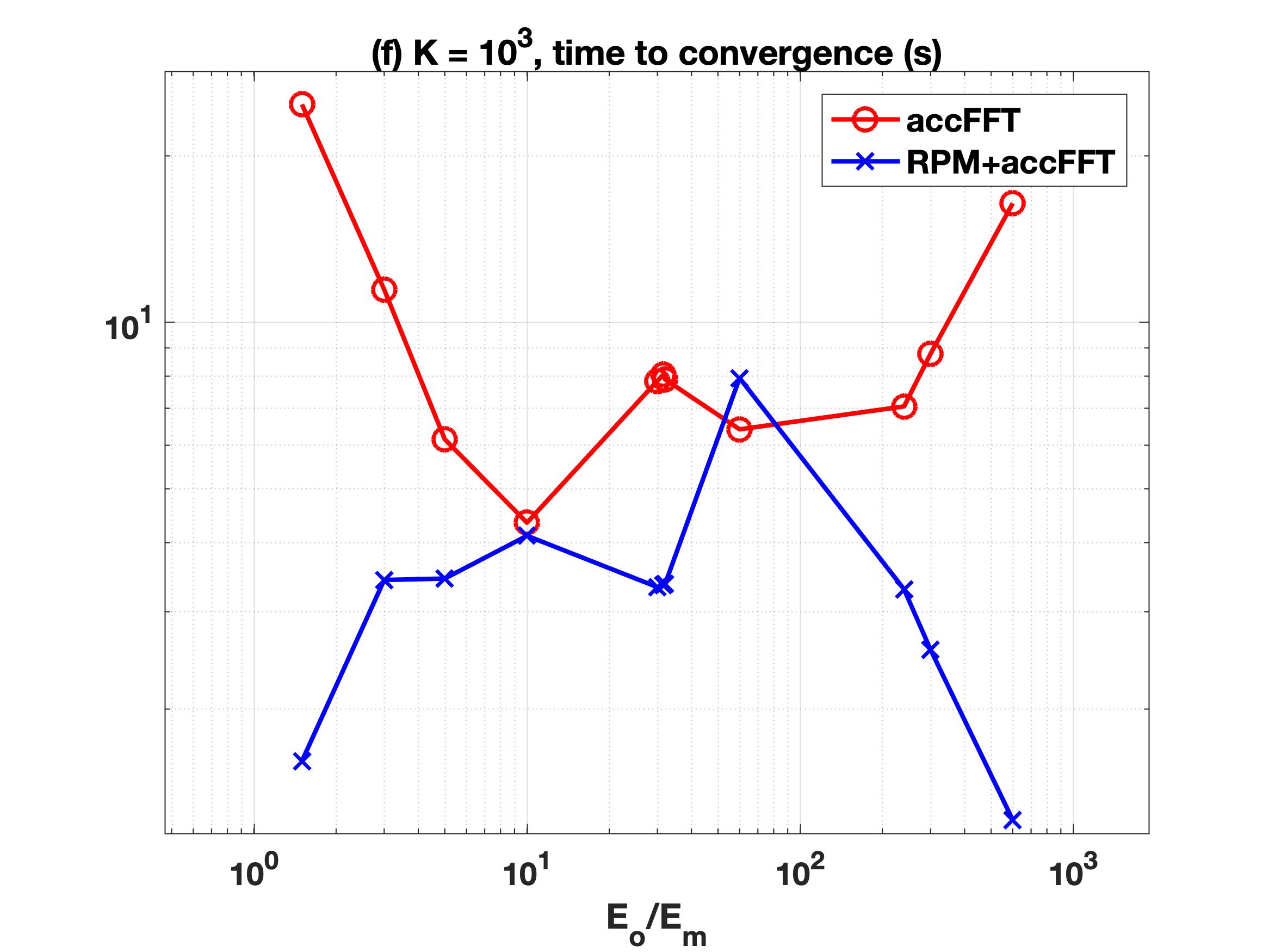}
                    \\
                    \includegraphics[width=80mm]{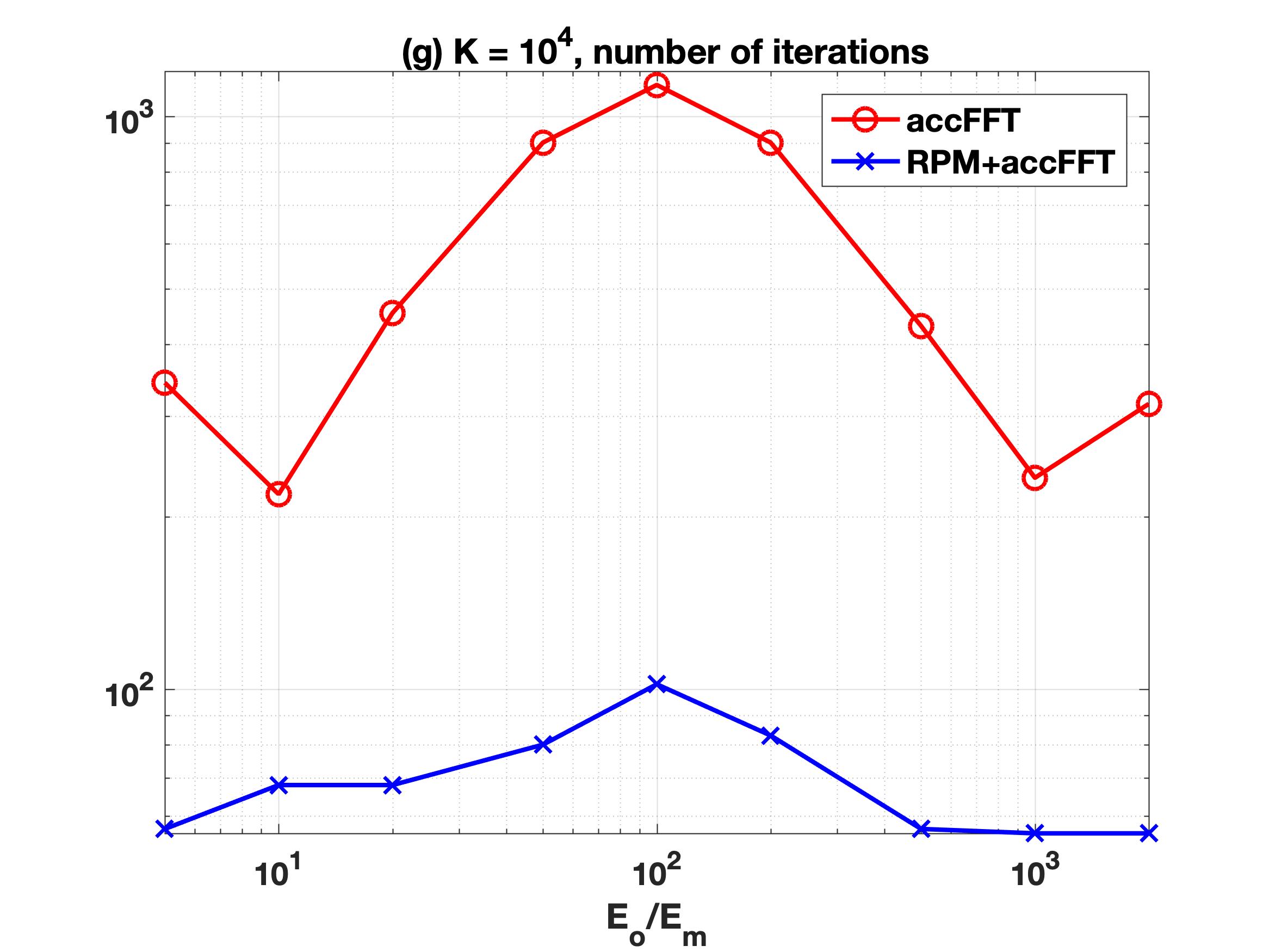}
                    \includegraphics[width=80mm]{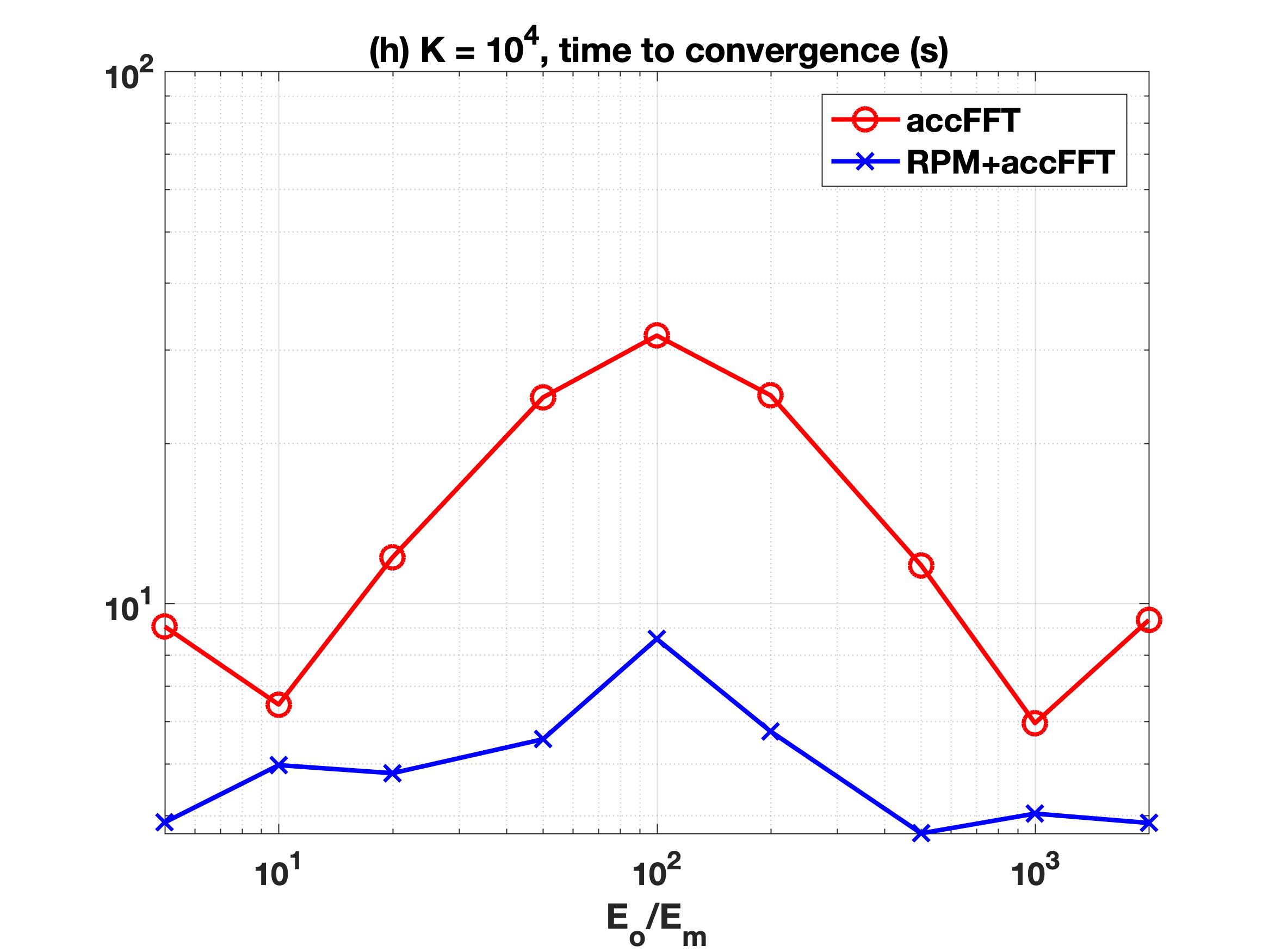}
					\caption{Number of iterations and times to convergence as a function of  reference medium modulus $E_o$.  We examine $K=10, 10^2, 10^3, 10^4$ in the first, second, third and fourth rows respectively.}
					\label{fig:accRPM}
				\end{center}
			\end{minipage}
		}
	\end{center}
\end{figure}

We next examine the sensitivity to the choice of reference medium for  Accelerated FFT and RPM-Accelerated FFT.
Fig. \ref{fig:accRPM} compares the time to convergence and number of iterations -- for a wide range of $K$ -- for the usual Accelerated FFT method against the RPM-Accelerated FFT method, for various choices of reference medium.

When K is small, the performance of the Accelerated FFT follows what has been observed in \cite{moulinec2014comparison}: namely, the best choice is at when $E_o / E_f  = \sqrt K$.
But this is not the case for $K = 10^3, 10^4$. Meanwhile, RPM-Accelerated FFT is much less affected by the different choices of the reference over all of the $K$s being tested. 
As pointed out in \cite{moulinec2014comparison}, the $E_o / E_f  = \sqrt K$ choice is valid when the precision is high ($10^{-10}$), which explains Fig. \ref{fig:accRPM} (e,f,g,h), as all tests here are computed with $10^{-4}$ precision which is more in line with standard practice.

We notice, particularly at high contrast, that the use of RPM in conjunction with Accelerated FFT makes the method much more robust with regard to the choice of reference medium.


\section{Concluding Remarks}



We have applied RPM \cite{shroff1993stabilization} to the problem of determining the stress and strain in the unit cell of a periodic linear elastic material.
Existing methods for this problem, e.g. Classical FFT \cite{moulinec1998numerical} and Accelerated FFT \cite{michel2001computational} methods, are fixed-point methods, and can have difficulty converging when the eigenvalues of the Jacobian lie outside the unit disk.
While Newton methods can recover the convergence, they are very expensive on the high-dimensional problems of interest.
RPM \cite{shroff1993stabilization} provides an elegant and efficient balance between fixed-point and Newton methods: it uses the fixed-point iterations to adaptively identify the unstable subspace that requires Newton, and performs Newton only on that subspace.
Fixed-point iterations can be performed on the complementary stable subspace.
For practical reasons of not changing existing algorithms and code, fixed-point is performed on the entire space, but is then projected to the complementary stable subspace.

A variational perspective, using analogies from electromagnetism, provides insight into the reason that the RPM decomposition can work well.
In particular, we expect that there are flat or unstable directions in the energy landscape in the context of certain formulations of heterogeneous linear elasticity, particularly when the stiffness tensor can vanish, as in the case of voids in materials.
The RPM-FFT method proposed in this paper exploits this structure to perform Newton iterations in directions along which the fixed-point would be unstable.

Our results show that the RPM-FFT is more efficient than the fixed-point methods as the contrast is increased.
Further, all of the fixed-point methods mentioned above require the use of a homogeneous reference medium as an important part of the method, but there is typically no systematic approach to selecting the reference medium.
We see that while the fixed-point methods can be sensitive to the choice of reference medium, the RPM-FFT method is extremely robust in terms of convergence for a wide range of choices for the reference medium.

We also note the feature of RPM that it can be easily used to ``wrap-around'' any fixed-point algorithm.
Consequently, while we have used it here for two specific instances of FFT-based methods, it can be easily combined with other formulations if they are found to be more efficient.

While we have applied RPM-FFT to the context of mechanical response, our overall approach is applicable to numerous linear problems with the same structure in homogenization \cite{milton2002theory}.
Further, while we have not examined nonlinear problems here, it is relatively straightforward to do this.
Briefly, most methods for nonlinear problems using the FFT consist of the repeated use of linear solves \cite{lebensohn2020spectral}, and the method presented can readily the other approaches in a modular way.

Finally, we note that while we have used Newton iterations for the space that is unstable under fixed-point iterations, it is possible to replace or combine Newton iterations with a conjugate gradient or other solver that could provide better performance \cite{vondvrejc2014fft,gelebart2013non,kabel2014efficient }.
Potential improvement can be achieved by combining the model order reduction techniques for FFT solvers as proposed in \cite{kochmann2019simple} with RPM. 
This provides a direction for future exploration.
In this context, we particularly highlight the recent interesting approach proposed in \cite{schneider2017fft}, where a different variational principle is shown to provide promising results.


\section*{Acknowledgments}

We thank Timothy Breitzman, Anthony Rollett, and Yi-Chung Shu for useful discussions.
We acknowledge financial support from ARO Numerical Analysis (W911NF-17-1-0084), ONR Applied and Computational Analysis (N00014-18-1-2528), NSF DMREF Program (1628994), NSF Mechanics of Materials and Structures (1635407), and ARO MURI Program (W911NF-19-1-0245).
We acknowledge the National Science Foundation for computing resources through the XSEDE program (TG-DMR120046) provided by Pittsburgh Supercomputing Center.
Kaushik Dayal acknowledges an appointment to the National Energy Technology Laboratory Faculty Research Participation Program sponsored by the U.S. Department of Energy and administered by the Oak Ridge Institute for Science and Education.

\section*{Availability of Codes}

The code developed and used for the calculations in this paper is available at 
\\ \url{https://github.com/KaushikDayalGroup/RPM-for-FFT-and-Elasticity}


\bibliographystyle{alpha}
\bibliography{rpm-refs}

\end{document}